%% file: main.tex
\documentclass[journal]{IEEEtran}
\usepackage{amsmath,amsfonts}
\usepackage{algorithmic}
\usepackage{algorithm}
\usepackage{array}
\usepackage[caption=false,font=normalsize,labelfont=sf,textfont=sf]{subfig}
\usepackage{textcomp}
\usepackage{stfloats}
\usepackage{url}
\usepackage{verbatim}
\usepackage{graphicx}
\usepackage{cite}
\usepackage{color}
\usepackage{booktabs}
\usepackage{multirow}
\usepackage{makecell}
\usepackage{threeparttable}
\usepackage{colortbl}
\usepackage{xcolor}
\begin{document}

	\title{
		BiGSeT: Binary Mask-Guided Separation Training for DNN-based Hyperspectral Anomaly Detection
	}
	
	\author{Haijun Liu,~\IEEEmembership{Member,~IEEE}, Xi Su, Xiangfei Shen, Lihui Chen and Xichuan Zhou,~\IEEEmembership{Senior Member,~IEEE}
		\thanks{This paper was supported in part by the National Natural Science Foundation of China under Grant 62001063, Grant 61971072 and Grant U2133211; in part by the Graduate Research and Innovation Foundation of Chongqing, China, under Grant CYB22068; and in part by the China Postdoctoral Science Foundation under Grant 2020M673135. Corresponding author: Xichuan Zhou.}
		\thanks{H. Liu, X. Su, X. Shen, L. Chen and X. Zhou are with the School of Microelectronics and Communication Engineering, Chongqing University, Chongqing 400044, China.}}
	
	\markboth{IEEE TRANSACTIONS ON IMAGE PROCESSING}%
	{Liu \MakeLowercase{\textit{et al.}}: BiGSeT for Hyperspectral Anomaly Detection}
	
	
	\maketitle
	
	\newcommand{\proposed}{BiGSeT}
	
	\begin{abstract}
		Hyperspectral anomaly detection (HAD) aims to recognize a minority of anomalies that are spectrally different from their surrounding background without prior knowledge. Deep neural networks (DNNs), including autoencoders (AEs), convolutional neural networks (CNNs) and vision transformers (ViTs), have shown remarkable performance in this field due to their powerful ability to model the complicated background. However, for reconstruction tasks, DNNs tend to incorporate both background and anomalies into the estimated background, which is referred to as the identical mapping problem (IMP) and leads to significantly decreased performance. To address this limitation, we propose a model-independent binary mask-guided separation training strategy for DNNs, named \proposed. Our method introduces a separation training loss based on a latent binary mask to separately constrain the background and anomalies in the estimated image. The background is preserved, while the potential anomalies are suppressed by using an efficient second-order Laplacian of Gaussian (LoG) operator, generating a pure background estimate. In order to maintain separability during training, we periodically update the mask using a robust proportion threshold estimated before the training. In our experiments, We adopt a  vanilla AE as the network to validate our training strategy on several real-world datasets. Our results show superior performance compared to some state-of-the-art methods. Specifically, we achieved a 90.67\% AUC score on the HyMap Cooke City dataset.
        Additionally, we applied our training strategy to other deep network structures, achieving improved detection performance compared to their original versions, demonstrating its effective transferability. The code of our method will be available at https://github.com/enter-i-username/BiGSeT.
	\end{abstract}


	\section{Introduction}
	
	\IEEEPARstart{I}{n} recent years, the research on hyperspectral imagery (HSI) has been gaining considerable attention in remote sensing due to its abundant spectral information for distinguishing materials \cite{ref_dis_material_1, ref_dis_material_2}. The advantage of the high spectral resolution makes it possible to apply in many fields like land cover classification \cite{ref_classification_1, ref_classification_2}, spectral unmixing \cite{ref_spectral_unmixing_1, ref_spectral_unmixing_2, ref_spectral_unmixing_3} and anomaly detection \cite{ref_anomaly_detection_1, ref_anomaly_detection_2, ref_anomaly_detection_3, ref_anomaly_detection_4, ref_anomaly_detection_5}. Among them, hyperspectral anomaly detection (HAD) is becoming one of the hotspots in remote sensing because of its essentiality for many applications \cite{ref_had_application_1, ref_had_application_2, ref_had_application_3} and also because of the challenging nature of the problem. The HAD task aims to identify an extremely small number of pixels that differ significantly in their spectral signature from the surrounding or overall background pixels, without relying on any prior information \cite{ref_no_prior_knowledge_1, ref_no_prior_knowledge_2}. The following three characteristics mainly cause the detection task to be difficult: 1) the absence of spectral information in advance for the desired target pixels; 2) the imbalance of samples between the anomalies and background in HSI; and 3) the complex diversity of the background, which leads to some pixels being erroneously recognized as abnormal.

    Due to the challenges posed by 1) and 2), existing methods primarily concentrate on unsupervised background modeling, which exposes abnormality by measuring the deviation degree of pixels from the background patterns directly learned from the HSI. They can be divided into three branches, including two traditional categories of statistical-based and representation-based methods, as well as deep learning-based methods. One of the early attempts is the statistical-based algorithm Reed-Xiaoli (RX) \cite{ref_rx}. This method models the background from a probability distribution perspective, assuming that the majority of background pixels follow a multivariate Gaussian distribution. It calculates the covariance and mean of the HSI as the background model, and measures the differences between each pixel and the model using the Mahalanobis distance. Following the RX, various methods have been proposed to improve the statistical background modeling. To utilize more local information, the local RX (L-RX) \cite{ref_local_rx} builds the background model in a window sliding manner where the window is centered around each pixel. The weighted RX (W-RX) \cite{ref_weighted_rx} provides a better background estimation by developing a weight assignment strategy for each pixel. Moreover, some methods \cite{ref_kernel_1, ref_kernel_2, ref_kernel_3} adopt the kernel tricks to fit more complex non-Gaussian distributions. However, in real-world hyperspectral scenes, ideal distributions of background are not always satisfied. 

    Some other efforts have been dedicated to the representation-based methods. They hold a weak assumption that in a local or global homogeneous scene, background pixels can be easily represented by other pixels in the HSI, while the anomalies cannot \cite{ref_representation_1, ref_representation_2, ref_representation_3}. Thus, a background model is first learned to represent each pixel, and the residual errors between the representation and the original HSI are used to measure the abnormality. Based on this assumption, a variety of representation-based approaches have been proposed. The collaborative representation-based detector (CRD) \cite{ref_crd} regards each background pixel as a linear combination of their neighboring pixels. To explore the global property of the background, the low-rank representation is introduced in \cite{ref_lrasr}. Furthermore, the sum-to-one and non-negativity constraints are utilized in the representation to improve the physical interpretability \cite{ref_more_constraints}. In addition to the background component, the potential anomalies are also represented by other pixels to enhance the model \cite{ref_pab_dc}. Although this group of methods can achieve high detection performance in monotonous and simple scenes, the representation may fail when the background contains more complex diversity, which limits their practical applications.

    Recently, the successful application of deep learning-based methods in remote sensing \cite{ref_dl_in_remote_sensing_1, ref_dl_in_remote_sensing_2, ref_dl_in_remote_sensing_3} has also shown powerful advantages in HAD \cite{ref_dl_in_had_1, ref_dl_in_had_2, ref_dl_in_had_3}. One of the popular methods among them is the autoencoder (AE), which can learn nonlinear and high-level features of an HSI in an unsupervised manner to handle complex background modeling. Generally, an AE is trained as a background reconstructor for a given HSI, and the anomalousness is measured using reconstruction errors. Early works appeared on sparse AEs \cite{ref_sparse_ae_1, ref_sparse_ae_2}, where researchers imposed constraints on the sparsity of AEs. To exploit more local spatial information, additional regularizations were incorporated. Lu \textit{et al}. \cite{ref_embedding_manifold_ae} utilized the embedding manifold of AEs to reflect the intrinsic structure of the HSI. Fan \textit{et al}. \cite{ref_rgae} proposed the graph regularization based on superpixel segmentation to preserve the local spatial consistency of the HSI. In addition to the simple AE architecture, other deep neural networks (DNNs) have also been extensively studied due to their ability to automatically extract deeper and more abstract features from the input data \cite{ref_dnn_1, ref_dnn_2, ref_dnn_3}.

    \begin{figure}[!t]
		\centering
		\subfloat[]{\includegraphics[scale=0.29]{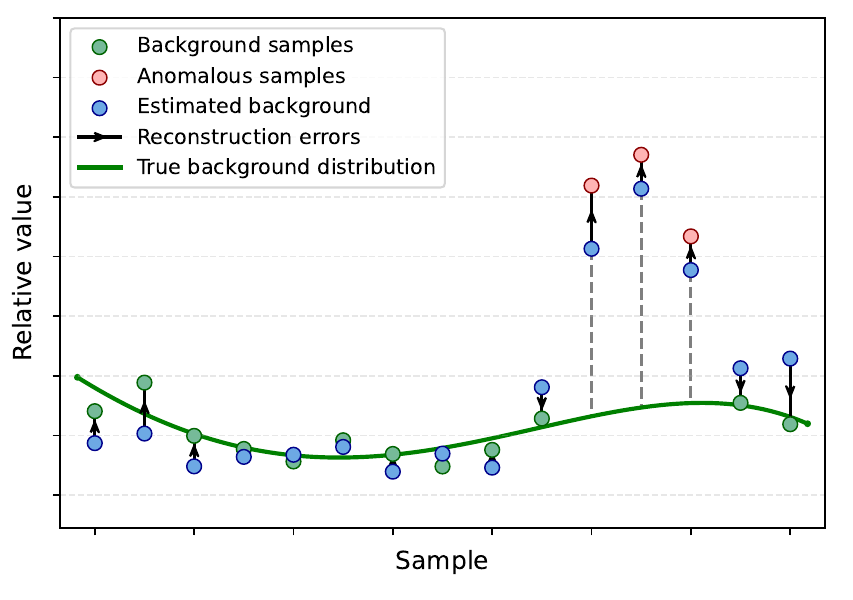}}
			\label{fig_imp_phenomenon}
		\hfill
		\subfloat[]{\includegraphics[scale=0.29]{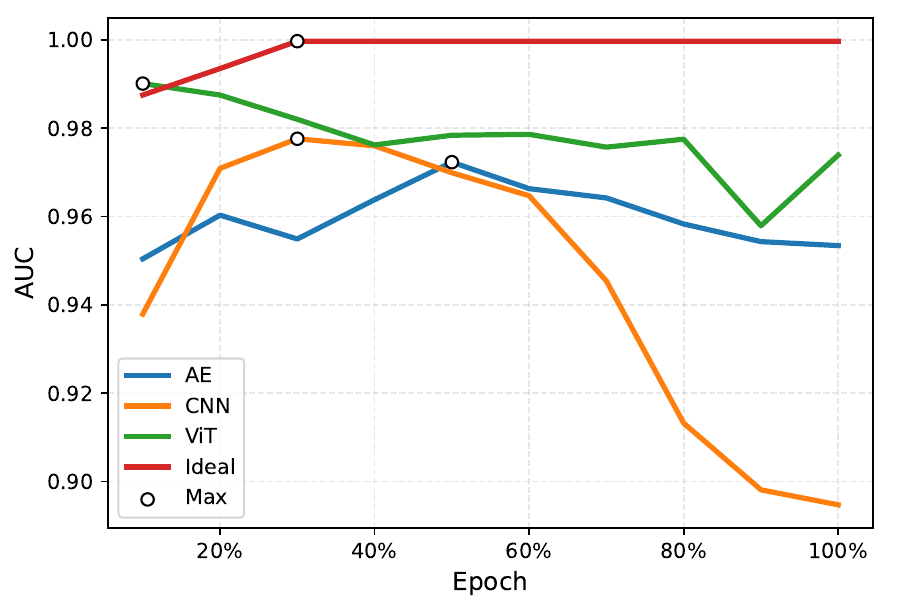}}
			\label{fig_imp_performance}
		\caption{Illustration of the IMP. (a) The DNN attempts to minimize reconstruction errors (arrowed black solid lines) between the estimated background (blue points) and the original HSI samples (green and red points), regardless of whether a sample belongs to the background or not. However, for those anomalous samples, the estimated background gets closer to the anomalies (red points), while simultaneously deviating from the true background distribution (green curve), resulting in inseparable reconstruction errors and decreased performance. (b) The AUC scores of the AE, CNN and ViT (blue, orange, and green curves, respectively) decline as the training progresses due to the IMP. In contrast, the ideal training process we expect (in red) maintains high performance throughout the training. }
		\label{fig_imp}
    \end{figure}

    Although reconstruction DNNs can handle the complex background modeling, a more difficult issue arises, the \textit{identical mapping} problem (IMP), also defined as the ``identical shortcut'' problem \cite{ref_identical_shortcut}. We expect a DNN to serve as a pure background reconstructor of the input HSI. However, with the progress of training, it inevitably involves the anomalies in the reconstructed image. This is because models favor learning all the information from input, including both background and anomalies simultaneously. As a consequence, the learned background reconstructor would obtain not only the background but also the anomalies together. Therefore, the background and anomaly pixels cannot be separated by the resulting reconstruction errors, leading to decreased performance. To clearly illustrate this phenomenon, we give a description of the IMP in Fig. \ref{fig_imp}. Note that this problem is widely prevalent in DNN structures such as AE, CNN and ViT, affecting the detection performance and hindering further research in DNN-based HAD.
 
    Some semi-supervised methods provide a perspective from dataset selection to address this problem \cite{ref_adversarial_framework, ref_ls3tnet}. They choose pure background pixels as the training dataset to reconstruct the original HSI, desiring more intensive separability of the reconstruction errors. However, if the training dataset is not properly cleansed and instead contains anomalies, models lack countermeasures to eliminate these bad samples during training. An alternative approach emphasizes model training \cite{ref_rgae, ref_auto_ad}, instead of dataset purification. Specifically, during the training stage, anomalies are continuously prevented from being reconstructed, and this process is thus referred to as ``anomaly suppression''. Until now, this category of methods has not been extensively studied yet. Also, they are currently only discussed on specific networks, which limits their potential applications in other DNNs. Therefore, the research on a general framework that can effectively suppress anomalous targets during training is of crucial importance and urgently needed.

    To address the IMP in DNN-based HAD, we propose a general binary mask-guided separation training framework in this article, named \proposed. In the process of model training, we explicitly separate the background and anomalies instead of treating them equally, which is the main cause of the IMP. To achieve this separation, we propose the utilization of a latent binary mask matrix that identifies potential anomalous and background pixels, thereby guiding the training process towards optimal performance. Based on this mask, we propose a separation training loss function that reconstructs the background while suppressing anomalies, thus facilitating the learning of pure backgrounds. Furthermore, for efficient suppression of anomalies, we employ the Laplacian of Gaussian (LoG) regularization, a second-order operator that mitigates the significant spatial variations of the anomalies during training. As we lack prior knowledge of anomalies, we generate the mask and regularly update it to ensure separability throughout the training process by binarizing reconstruction errors. To obtain a robust mask, we employ the statistics of the HSI to estimate the proportion threshold for the mask through a distribution-adjusted unimodal thresholding algorithm prior to training. Our \proposed~method is model-independent, making it applicable to various network structures. We conduct experiments to validate this advantage of our method, opening up new perspectives for its application in DNN-based HAD.
	
    The main contributions are listed as follows.
    \begin{enumerate}
        \item{We propose a general separation training framework, \proposed, for DNN-based HAD to prevent the learning of \textit{identical mappings}. \proposed~is based on a periodically updated binary mask matrix, and it separately considers the background and potential anomalies, preserving the former while suppressing the latter. }
        \item{We propose an efficient regularization for anomaly suppression using the LoG operator, which estimates a pure background by alleviating large spatial variations of the anomalies. }
        \item{We utilize the statistics of the HSI to estimate the proportion threshold through the distribution-adjusted unimodal thresholding algorithm to yield a more robust mask. }
        \item{Our BiGSeT method outperforms other state-of-the-art HAD methods on the ABU benchmark dataset, demonstrating its adaptability across various scenes. Additionally, BiGSeT obtains an AUC score of 90.67\% on the large HyMap Cooke City dataset, which features a more complex background.}
    \end{enumerate}

	\section{Related Work}
 
	In this section, we will discuss DNN-based methods for HAD. Firstly, we will provide a brief introduction to some deep neural networks. Then, we will focus on some solutions to the IMP.
	
	\subsection{Deep Neural Networks for HAD}
	
	The most widely used networks are the convolutional neural networks (CNNs). Hosseiny \textit{et al}. \cite{ref_convs} used 1-D and 2-D stacked CNNs for extraction of deep and nonlinear relations. Cao \textit{et al}. \cite{ref_msbrnet} built up their network by cascading a low-rank module and a multiscale module. Wang \textit{et al}. \cite{ref_auto_ad} achieved autonomous detection using a fully convolutional AE with skip connections. Wang \textit{et al}. \cite{ref_ls3tnet} proposed a two-stream network that integrates local spatial-spectral information. The upstream network focuses on extraction of spatial features while the downstream network learns the distribution of the background, and the final detection map is integrated from the two streams. Very recently, Xiao \textit{et al}. \cite{ref_s2dwmtrans} for the first time introduced the vision transformer (ViT) structure into the field of HAD and found its effectiveness in detection tasks.
	
	\subsection{Background Purification before Training}
	
	The basic principle of the background purification methods is to remove potential anomalies and select pure background pixels as training samples in an unsupervised manner before training the network. An adversarial learning framework proposed in \cite{ref_adversarial_framework} utilized the density-based spatial clustering of applications with noise (DBSCAN) \cite{ref_dbscan} after reducing the dimensionality of the original HSI, to reject low-density anomalies and noise. Based on the DBSCAN algorithm, Li \textit{et al}. \cite{ref_sparse_had} searched their background data. They fed the data into a sparse coding-inspired generative adversarial network (GAN), and trained the network in an end-to-end manner. A simpler method was proposed in \cite{ref_ls3tnet}, which assumes that the anomalies are larger on the Mahalanobis distance space, and the largest part is removed from the training samples. In addition to pixel-level background extraction, a superpixel-level method was also proposed in \cite{ref_superpixel_background_extraction} to exploit more spatial information. The pixels were first segmented and clustered, and then the rare clusters were recognized and removed using connected domain searching.
	
	This category of methods performs background purification and network training independently, which may still pose a risk of incomplete removal of anomalies, potentially allowing DNNs to learn from remnants of anomalies in the training stage.
	
	\subsection{Anomaly Suppression during Training}
	
	The other branch usually takes all pixels in an image as training samples and employs anomaly suppression strategies to reduce the impact of anomalies during the training stage. In \cite{ref_rgae}, the $l_{2, 1}$-norm is utilized, from the perspective of gradients, to mitigate the sensitivity of the model to abnormal targets and noise. Recently, in \cite{ref_auto_ad} and \cite{ref_s2dwmtrans}, two adaptive-weighted (AW) strategies along with their deep neural networks were proposed to suppress the anomalies during training, respectively. These AW methods generate a converse weight map calculated based on the reconstruction errors, which is incorporated into the loss function during the training process to suppress potential anomalies. Another method \cite{ref_gaed} introduced a guided module embedded in the AE network, which directs the training process towards pure background reconstruction. 
	
	Although these methods allow for continuous adjustment of the anomaly suppression during training, they still have some limitations. For instance, the $l_{2, 1}$-norm method does not incorporate direct spatial information for DNNs.  The AW methods may only weakly suppress anomalies using a soft weight map, which could potentially result in inadequate separation of anomalies from the background during the training stage. Moreover, they are only tested on specific networks, and the transferability to other networks remains to be explored.

	\section{Proposed Method}
	
	\subsection{Observation and Motivation}
	
    As shown in Fig. \ref{fig_imp} (b), DNNs get decreased performance due to the IMP as training progresses. This occurs because the signatures of the background and anomalies are not explicitly distinguished during training, but each element of the reconstructed image is treated equally \cite{ref_rgae}. As a result, the identical mapping is inevitably learned, even if the patterns of anomalies are relatively hard to dig up \cite{ref_auto_ad}. Consequently, the reconstruction errors become inseparable as they converge to zero at a very similar rate. 
 
    In comparison, the ideal training process depicted in Fig. \ref{fig_imp} (b) exhibits a steadily increasing detection performance that reaches its maximum and remains stable at a plateau throughout the training stage. Since the performance does not drop, the termination of the unsupervised training can be relaxed, thus we can get higher and more robust results. To this end, two conditions need to be satisfied: 1) we separately consider the anomalies and background in the estimated background image, suppressing the anomaly part while reconstructing the background part; 2) the separation of the two parts reaches a dynamic equilibrium such that the detection performance can be maintained at a high level. Notably, the separation itself is independent of the model's form, which allows us to apply this separation strategy to any reconstructor.

    Therefore, we design a model-independent training framework to help DNNs overcome the IMP, based on these two conditions.

	\subsection{Overview of the Proposed Method}
	
	Given a raw HSI $\textbf{X} \in \mathbb{R} ^{H \times W \times L}$, where $H$, $W$ and $L$ respectively denote the height, width and spectral bands, our goal is to generate a background image $\hat{\textbf{X}} \in \mathbb{R} ^{H \times W \times L}$, and the reconstruction error matrix $\textbf{R} \in \mathbb{R}_{+}^{H \times W}$ is obtained by
	\begin{equation}
	\label{eq_error_map}
	\textbf{R}_{i, j} = \Vert \hat{\textbf{X}}_{i, j, :} - \textbf{X}_{i, j, :} \Vert_2^2,
	\end{equation}
	where $\textbf{R}_{i, j}$ represents the non-negative scalar error at position $(i, j)$, and $\hat{\textbf{X}}_{i, j, :}$ and $\textbf{X}_{i, j, :}$ are the spectral vectors, respectively.
	
	\begin{figure*}[!t]
		\centering
		\includegraphics[scale=0.7]{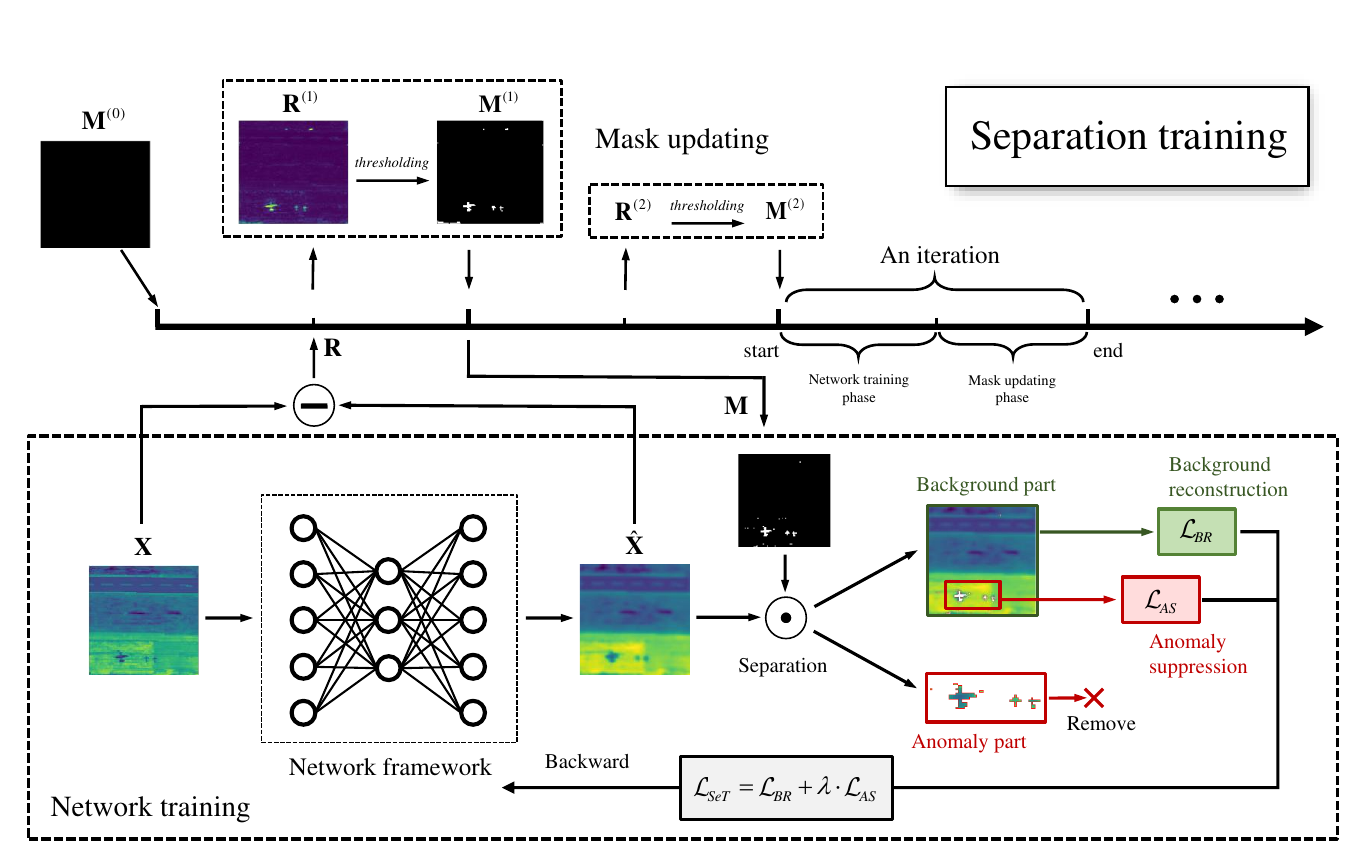}
		\caption{Flowchart of our proposed method. Our training process consists of multiple iterations, with a binary mask that is periodically updated throughout the process. In each iteration, we first train the network and obtain the reconstruction errors. Then, we update the mask. In the network training phase, we divide the reconstructed image into background and potential anomaly parts using the mask, and apply separate constraints to each part. In the mask updating phase, the mask is replaced with thresholded reconstruction errors.} 
		\label{fig_flowchart}
	\end{figure*}
	
    We introduce a binary mask $\textbf{M}$, where ones indicate the potential anomalies and zeros are the background, into our training to separately constrain the anomalies and background, and periodically update the mask to maintain the separability. 
 
    The flowchart of the proposed method is shown in Fig. \ref{fig_flowchart}. The whole training consists of multiple iterations. In each iteration, 1) we first train the network and yield $\textbf{R}$ (the network tarining phase), 2) subsequently update $\textbf{M}$ by binarizing $\textbf{R}$ (the mask updating phase), and then 3) pass it down to the next iteration. 
	
	In the network training phase, the estimated image $\hat{\textbf{X}}$ is divided into two mutually exclusive parts by $\textbf{M}$, i.e. the anomaly part and background part. The total training loss is given below to separately train the two parts:
	\begin{equation}
	\label{eq_total_loss}
	\mathcal{L}_{SeT} = \mathcal{L}_{BR} + \lambda \cdot \mathcal{L}_{AS},
	\end{equation}
	where the loss function $\mathcal{L}_{BR}$ reconstructs the background part and $\mathcal{L}_{AS}$ suppresses the anomaly part; $\lambda$ is a tradeoff hyperparameter.
	
	In the mask updating phase, a new mask is obtained to replace the one from the last iteration by binarizing $\textbf{R}$ via a proportion threshold, which is estimated by the unimodal thresholding algorithm \cite{ref_unimodal_thresholding} as an initial parameter before training.
	
	\subsection{Separation Training}
	
	To address the IMP, we introduce a binary mask $\textbf{M}$ which explicitly distinguishes the potential anomalies from the background. The background needs to be reconstructed during training as close as possible to that of the input image, whereas the anomalies need suppression from the original HSI. We herein give the form of the separation training loss function:
	\begin{equation}
	\label{eq_BR_loss}
	\mathcal{L}_{BR} = \frac{1}{S(\Bar{\textbf{M}})} \Vert (\hat{\textbf{X}} - \textbf{X}) \odot \Bar{\textbf{M}} \Vert_F^2,
	\end{equation}
 
	\begin{equation}
	\label{eq_AS_loss}
	\mathcal{L}_{AS} = \frac{1}{S(\textbf{M}) + \epsilon} \mathcal{R}(\textbf{M}, \hat{\textbf{X}}),
	\end{equation}
    where $\odot$ represents the element-wise multiplication over spatial dimensions and $\Bar{\textbf{M}}$ is the Boolean \textit{not} operation on \textbf{M}; $S(\cdot)$ denotes the number of ones (we have $S(\Bar{\textbf{M}}) = H \times W$ and $S(\textbf{M}) = 0$ if $\textbf{M} = \textbf{0}$, and a very small constant $\epsilon$ is added to avoid division by zero); $\mathcal{R}(\textbf{M}, \hat{\textbf{X}})$ indicates the regularization for suppressing the anomalies selected by $\textbf{M}$ in the estimated image $\hat{\textbf{X}}$. Equation (\ref{eq_BR_loss}) describes the approximation of the estimated image to the original image in the background part. However, we do not expect to reconstruct the anomaly part, but utilize the distribution of the background to predict it. Therefore, the regularization term $\mathcal{R}(\textbf{M}, \hat{\textbf{X}})$ needs to learn the patterns of $\hat{\textbf{X}} \odot \Bar{\textbf{M}}$, and simultaneously generate background-like spectra at position $\textbf{M}$.
 
	\begin{figure}[!t]
		\centering
		\subfloat[]{\includegraphics[scale=0.45]{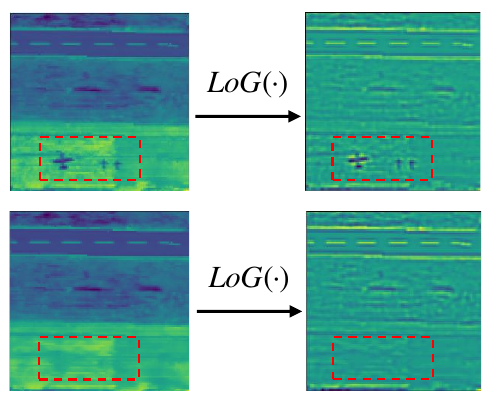}
			\label{fig_LoG_a}}
		\hfill
		\subfloat[]{\includegraphics[scale=0.45]{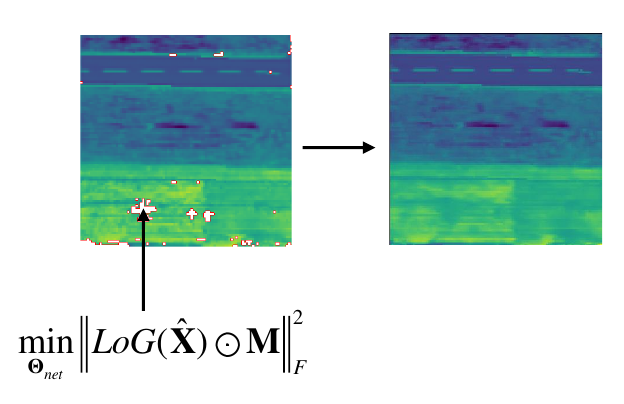}
			\label{fig_LoG_b}}
		\caption{Anomaly suppression using the LoG operator. (a) Edges of anomalies are obvious because the anomalous pixel groups have significant differences from their neighboring background, whereas these edges do not exsist in the anomaly-free background. (b) Filling in the blanks by minimizing LoG of the anomaly part such that the edge effects are alleviated and the pure background is estimated.}
		\label{fig_LoG}
	\end{figure}
	
    \textbf{Anomaly Suppression with LoG Operator.} To suppress the anomalies, or equivalently learn a pure background, we remove the selected anomalies by $\textbf{M}$, and fill in the blanks using the information from their neighboring background pixels. In this article, we adopt the well-known edge detection method Laplacian of Gaussian (LoG) to estimate the blanks. By imposing LoG of a whole blank region to be zero, there will not be large variations like edges in the region, achieving a smooth and natural transition from the boundary to the interior, as shown in Fig. \ref{fig_LoG}. Thus, the region is completely estimated from the pure background distribution and no anomaly remains. For more efficient calculations, we take a $\text{5} \times \text{5}$ LoG template \textbf{L} to convolve with the image $\hat{\textbf{X}}$ for each band:
	\begin{equation}
	\label{eq_LoG_template}
	\textbf{L} = 
	\begin{bmatrix} 
	-2 & -4 & -4 & -4 & -2 \\
	-4 &  0 &  8 &  0 & -4 \\
	-4 &  8 & 24 &  8 & -4 \\
	-4 &  0 &  8 &  0 & -4 \\
	-2 & -4 & -4 & -4 & -2
	\end{bmatrix},
	\end{equation}
	\begin{equation}
	\label{eq_LoG_operation}
	LoG(\hat{\textbf{X}}) = \textbf{L} \ast \hat{\textbf{X}},
	\end{equation}
	where $\ast$ is the 2D convolution operator. Before the LoG operation, we use a reflection padding of 2 to keep the spatial sizes consistent. The regularization for anomaly suppression can be then expressed as the following:
	\begin{equation}
	\label{eq_AS_regularization}
	\mathcal{R}(\textbf{M}, \hat{\textbf{X}}) = \Vert LoG(\hat{\textbf{X}}) \odot \textbf{M} \Vert_F^2 .
	\end{equation} 
	
	\subsection{Mask Updating}
	
	If the separation mask accurately reveals where the anomalous pixels locate, the model will never fall into the IMP, because the anomalies can be completely removed. However, a question arises: how do we obtain the mask if any prior knowledge is unavailable? Apparently, since the mask is an unobservable variable, we can only estimate it from the HSI itself.
	
	Suppose the detection is divided into two stages. The model is pre-trained and coarsely searches the potential anomalies in the first stage, and is fine-tuned to get a more precise result in the next stage. Specifically, let $\textbf{M}$ be $\textbf{0}$, which means there is no prior location knowledge for the anomalies, and we train the model with only the background reconstruction loss $\mathcal{L}_{BR}$ for a relatively small number of epochs (e.g. 150 epochs in this paper). We calculate the reconstruction error map $\textbf{R}$ as a coarse detection result, where larger values are likely to be the anomalies. Then the values in $\textbf{R}$ are sorted in ascending order, and listed into a vector $\tilde{\textbf{r}} = [\tilde{r}_1, \tilde{r}_2, \dots, \tilde{r}_{HW}]^T \in \mathbb{R}_{+}^{HW}$. We consider the first $\tau \cdot HW$ values in $\tilde{\textbf{r}}$ as the background part, where $\tau \in (0, 1]$ is a proportion threshold. $\textbf{M}$ is therefore updated for the next detection stage by
	\begin{equation}
	\label{eq_M}
	\textbf{M}_{i, j} = 
	\begin{cases}
	1, & \text{if } \textbf{R}_{i, j} > \tilde{r}_{ceil(\tau \cdot HW)}, \\
	0, & \text{otherwise},
	\end{cases}
	\end{equation}
	where the index $ceil(\tau \cdot HW)$ means the minimum integer no less than $\tau \cdot HW$. 
	
	\begin{figure}[!t]
		\centering
		\includegraphics[scale=0.46]{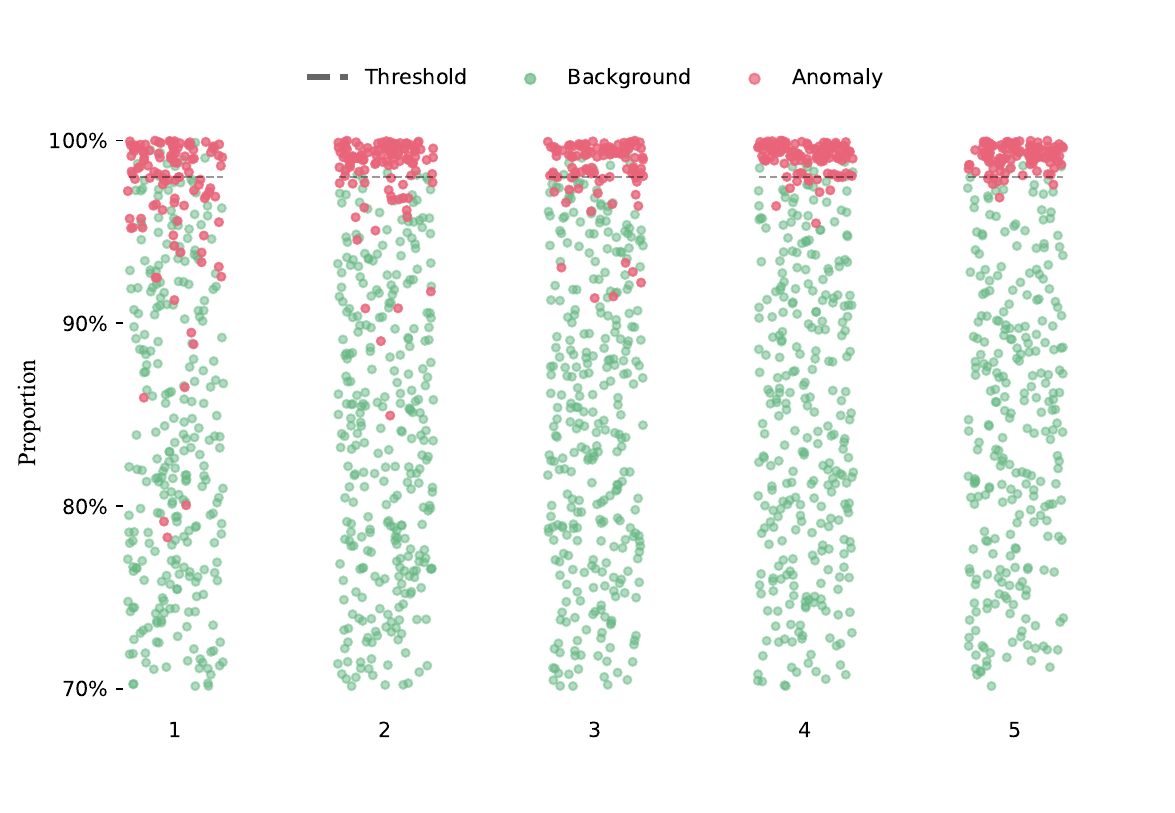}
		\caption{Training process with the periodic mask updating strategy. We sort the values in the detection map of 5 iterations and show the first 30\% of these data points. The anomalous points float up to the top and finally reach an equilibrium around the threshold line with the background points. }
		\label{fig_equilibrium}
	\end{figure}
	
	However, it must be noted that $\textbf{M}$ from the first coarse search may not be an accurate locator for the anomalies to be suppressed. If we perform a long-term training based on this mask in the first stage, the two parts will not converge to a well-separated status. Therefore, $\textbf{M}$ needs to be periodically updated throughout the training process to avoid the model falling into local minima. We extend the two-stage detection to totally $K$ iterations. In the $m$th iteration, the model is first trained with $\textbf{M}^{(m-1)}$ from the last iteration, and then the mask is updated to $\textbf{M}^{(m)}$ to pass down to the next stage (we define $\textbf{M}^{(0)} = \textbf{0}$). To further purify the background reconstruction in $m$th stage, we input the network with the selected background part $\textbf{X} \odot \Bar{\textbf{M}}^{(m-1)}$ instead of the whole image $\textbf{X}$, thus the impact of potential anomalies can be reduced. 
	
	Fig. \ref{fig_equilibrium} visually demonstrates the training process with the periodic mask updating strategy. According to the defination of $\textbf{M}$, we know that all the points above the threshold line are marked by $\textbf{M}$ as the potential anomalies. Due to the constraint of the mask updating strategy, the true anomalies eventually reach an equilibrium around $\tau$, and will never dive to the bottom. Therefore the separability is maintained.
	
	
	\textbf{Threshold Estimation.} Another essential step is to obtain the proportion threshold $\tau$ as an initial parameter to update the mask. This paper utilizes the global statistical information of the HSI to estimate an appropriate value of $\tau$ before training. 
	
\begin{figure}[!t]
		\centering
		\includegraphics[scale=0.387]{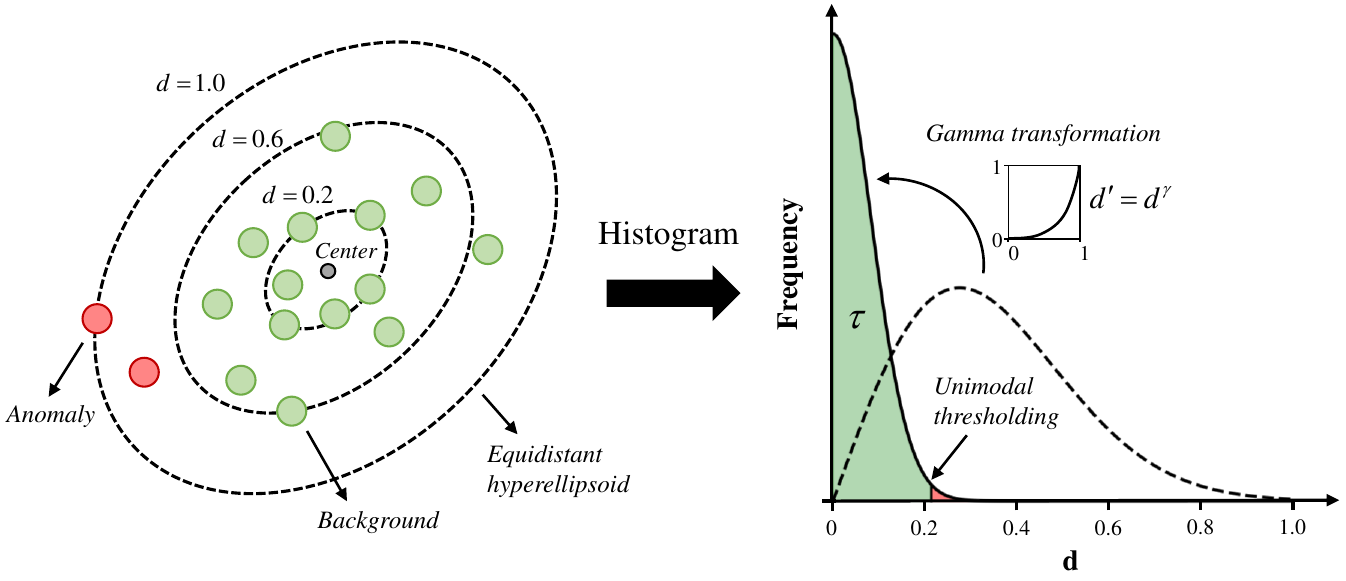}
		\caption{Threshold estimation. The relative Mahalanobis distance map is first calculated, and its statistical information is utilized to estimate the proportion threshold $\tau$ with the unimodal thresholding algorithm.}
		\label{fig_thresholding}
	\end{figure}

	To give a clearer explanation, we illustrate the thresholding process in Fig. \ref{fig_thresholding}. The Mahalanobis distance is first calculated to project the high-dimensional HSI onto a 1D relative distance space $d \in [0, 1]$. The majority of the points concentrated around the center are the background pixels, while others far from the center are considered as the outliers. The point density goes down with the increase of distance, hence the distribution of $d$ displays irregular unimodality, which is shown as the dashed curve on the right side in Fig. \ref{fig_thresholding}. To handle unimodal thresholding, we employ a simple algorithm \cite{ref_unimodal_thresholding} that identifies a corner point where the dominant population (background samples) changes to the minority (anomalous samples). Before we use the unimodal thresholding algorithm to separate the two parts, the distribution is adjusted by gamma transformation $d' = d^{\gamma}, \gamma \ge 1$ to the solid curve to make the peak more dense in the region close to 0, such that the density separation is more robust. It is worth noting that we do not directly use the threshold estimated by \cite{ref_unimodal_thresholding} but the corresponding proportion threshold $\tau$, which is simply obtained by calculating the area on the left side of the separation point (green part in Fig. \ref{fig_thresholding}). This is because the proportion is not dependent on distributions derived from any transformation, but rather reflects the inherent characteristics of the separation of the HSI itself. 

        \begin{algorithm}[t]
        \caption{Algorithm for training the network.}
        \label{alg_1}
        \begin{algorithmic}[1]
            \renewcommand{\algorithmicrequire}{\textbf{Input:}}
            \REQUIRE
            HSI data \textbf{X}, 
            tradeoff hyperparameter $\lambda > 0$,
            number of training iterations $K \geq 1$,
            parameter $\gamma \ge 1$ for gamma transformation.
            \renewcommand{\algorithmicrequire}{\textbf{Initialize:}}
            \REQUIRE
            $\textbf{M}^{(0)} = \textbf{0}$,
            obtain $\tau$ via unimodal thresholding.
            \FOR{$m=1:K$}
            \FOR{$n=1:150$}
            \STATE Feed the network with $\textbf{X} \odot \Bar{\textbf{M}}^{(m-1)}$ as input and yield $\hat{\textbf{X}}^{(m)}$;
            \STATE Calculate the loss using (\ref{eq_total_loss});
            \STATE Do backward propagation;
            \ENDFOR
            \STATE Calculate $\textbf{R}^{(m)}$ using (\ref{eq_error_map});
            \STATE Calculate $\textbf{M}^{(m)}$ using (\ref{eq_M});
            \ENDFOR
            \renewcommand{\algorithmicensure}{\textbf{Output:}}
            \ENSURE
            Detection map $\textbf{R}^{(K)}$.
        \end{algorithmic}
    \end{algorithm}

	The separation training strategy prevents reconstructing the anomalies, thus it can maintain a high detection performance without decreasing. The complete training algorithm is summarized in Algorithm \ref{alg_1}.
	
	\section{Experiments}
	
	In this section, we conduct a series of experiments to exhibit the superiority of our proposed method. All experiments have been carried out on a PC with an Intel Core i7-11800H CPU at 2.30 GHz with 16 GB RAM, and an NVIDIA 3070 GPU with 8GB of memory. We have implemented the algorithms and run the code in Python 3.8 and PyTorch 1.11 or MATLAB 2021b environments.
	
	\subsection{Experimental settings}
	
	\textbf{Datasets.} We evaluate the detection performance of our methods using the Airport-Beach-Urban (ABU) benchmark dataset series\footnote{Available at http://xudongkang.weebly.com/}, which comprises a total of 13 images, including 4 airport scenes, 4 beach scenes, and 5 urban scenes. Most of the datasets were captured using the Airborne Visible/Infrared Imaging Spectrometer (AVIRIS) sensor, with the exception of the fourth image in the beach series (named Beach IV), which was acquired using the Reflective Optics System Imaging Spectrometer (ROSIS-03) sensor in Pavia. The scenes are mostly 100 $\times$ 100 in spatial dimensions, except for Beach I and Beach IV, which are 150 $\times$ 150. The location labels for the target objects were created using the Environment for Visualizing Images (ENVI) software tool.
	
	Another dataset used for evaluation is the HyMap Cooke City\footnote{Available at http://dirsapps.cis.rit.edu/blindtest/}, which was collected by the HyMap sensor over the area surrounding Cooke City, Montana, USA. This image has larger dimensions of 280 $\times$ 800 and more complex background than ABU datasets. The objects of interest in this scene are small fabric panels and civilian vehicles.

	\begin{figure}[!t]
		\centering
		\subfloat[]{\includegraphics[scale=0.33]{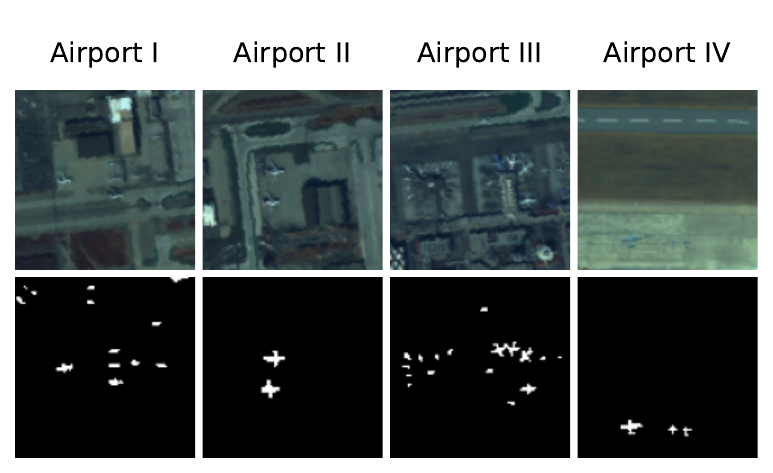}%
			\label{fig_airport}}
		\hfil\hfil
		\subfloat[]{\includegraphics[scale=0.33]{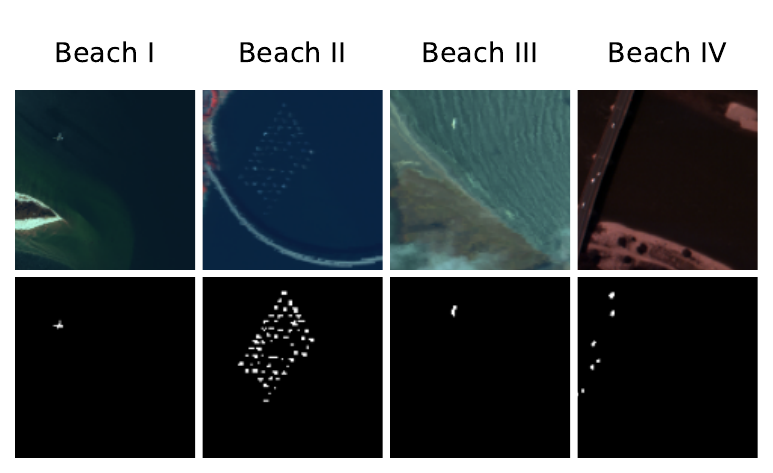}%
			\label{fig_beach}}
		\hfil\hfil
		\subfloat[]{\includegraphics[scale=0.33]{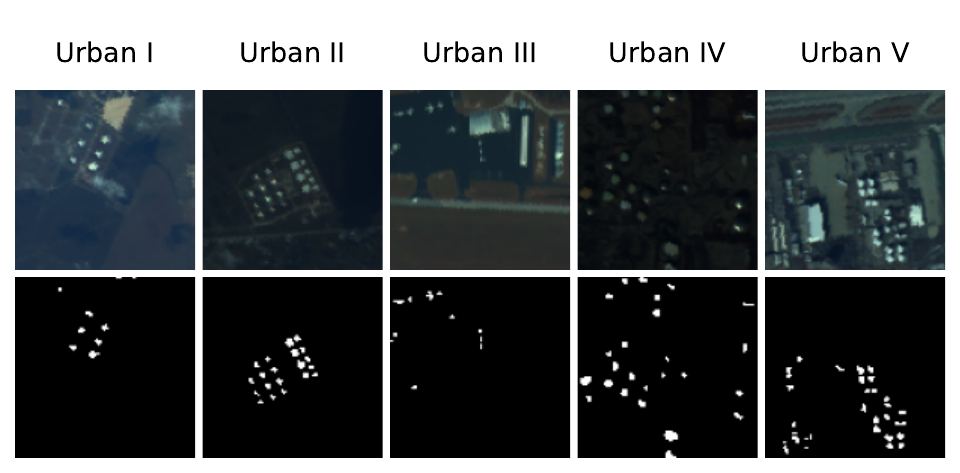}%
			\label{fig_urban}}
		\hfil\hfil
		\subfloat[]{\includegraphics[scale=0.33]{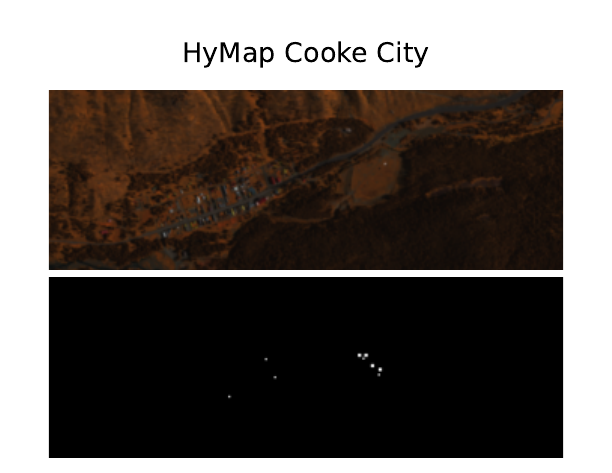}%
			\label{fig_hymap}}
		\caption{Pseudocolor images (upper row) and corresponding ground truth maps (lower row). (a) Airport. (b) Beach. (c) Urban. (d) HyMap Cooke City.}
		\label{fig_datasts}
	\end{figure}
	
	\begin{table}[!t]
		\caption{
			Description of the Datasets Used in This Paper
			\label{tab_dataset_description}
		}
		\renewcommand{\arraystretch}{1.3}
		\tabcolsep=0.2cm
		\centering
		\begin{tabular}{c c | c c c c}
			\toprule
			\multicolumn{2}{c|}{\textbf{Dataset}} & 
			\textbf{Location} & 
			\textbf{Resolution} & 
			\textbf{Bands} & 
			\makecell{\textbf{Target} \\ \textbf{pixels}} \\
			
			\hline
			\multirow{4}{*}{\textbf{Airport}} & \multicolumn{1}{|c|}{I}    & Los Angeles & 7.1 m   & 205 & 144         \\
			& \multicolumn{1}{|c|}{II}   & Los Angeles & 7.1 m   & 205 & 87 \\ 
			& \multicolumn{1}{|c|}{III}  & Los Angeles & 7.1 m   & 205 & 170        \\ 
			& \multicolumn{1}{|c|}{IV}   & Gulfport & 3.4 m   & 191 & 60        \\
			
			\hline\hline
			
			\multirow{4}{*}{\textbf{Beach}}   & \multicolumn{1}{|c|}{I}    & Cat Island & 17.2 m   & 188 & 19         \\
			& \multicolumn{1}{|c|}{II}   & San Diego & 7.5 m   & 193 & 202 \\ 
			& \multicolumn{1}{|c|}{III}  & Bay Champagne & 4.4 m   & 188 & 11        \\ 
			& \multicolumn{1}{|c|}{IV}   & Pavia & 1.3 m & 102 & 68        \\
			
			\hline\hline
			
			\multirow{5}{*}{\textbf{Urban}}   & \multicolumn{1}{|c|}{I}    & Texas Coast & 17.2 m   & 204 & 67         \\
			& \multicolumn{1}{|c|}{II}   & Texas Coast & 17.2 m   & 207 & 155 \\ 
			& \multicolumn{1}{|c|}{III}  & Gainesville & 3.5 m   & 191 & 52        \\ 
			& \multicolumn{1}{|c|}{IV}   & Los Angeles & 7.1 m   & 205 & 272        \\
			& \multicolumn{1}{|c|}{V}    & Los Angeles & 7.1 m   & 205 & 232        \\
			
			\hline\hline
			
			
			\multicolumn{2}{c|}{\multirow{2}{*}{\makecell{\textbf{HyMap}\\\textbf{Cooke City}}}} & \multirow{2}{*}{Cooke City} & \multirow{2}{*}{3 m} & \multirow{2}{*}{126} & \multirow{2}{*}{145} \\
			&&&&&\\
			

			\bottomrule
		\end{tabular}
	\end{table}
	
	The pseudocolor images and corresponding ground truth maps of the above datasets are displayed in Fig. \ref{fig_datasts}. We summarize the characteristics of these datasets, including the captured location, spatial resolution, spectral bands, and target pixels in Table \ref{tab_dataset_description} for easy reference.
	
	\textbf{Metrics.}
	In our experiments, the receiver operating characteristic (ROC) curve is used to evaluate the detection performance. Meanwhile, the area under curve (AUC) scores of $(P_d, P_f)$ are calculated as a quantitative evaluation. A method with a higher detection performance has an ROC curve located near the top-left corner which results in an AUC value closer to 1.
	
	\subsection{Performance Comparison}
	
	\subsubsection{Competitive Methods}
	
	Some competitive methods are selected as benchmarks for performance comparison. They are the statistical-based methods RX \cite{ref_rx}, W-RX \cite{ref_weighted_rx}, the representation-based methods LRASR \cite{ref_lrasr}, PAB-DC \cite{ref_pab_dc}, and four deep learning-based methods RGAE \cite{ref_rgae}, Auto-AD \cite{ref_auto_ad}, MSBRNet \cite{ref_msbrnet}, S2DWMTrans \cite{ref_s2dwmtrans}.
	
	For these deep learning methods, the RGAE trains a simple AE network with only one single hidden layer; Auto-AD and MSBRNet build up their networks with multiple convolutional layers; S2DWMTrans for the first time introduces the ViT structure into the field of HAD. 
 
    In this experiment, we only adopt a vanilla AE network that contains one linear layer with 100 hidden units, a Rectified Linear Unit (ReLU) nonlinear function, and an output linear layer. We use the ADAM optimizer \cite{ref_adam} to perform the backward propagation. 
	
	\subsubsection{Detection Results}
	
	\begin{table*}[!t]
		\caption{
			AUC Scores (Upper Row) and Time (Lower Row, in Seconds) for Different Methods on All Datasets.
			\label{tab_performance_comparison}
		}
		\renewcommand{\arraystretch}{1.5}
		\tabcolsep=0.2cm
		\centering
		\begin{threeparttable}
			\begin{tabular}{c | c c c c | c c c c c}
				\toprule
				\textbf{Dataset}\tnote{1} & 
				\textbf{RX}\tnote{2}  & 
				\textbf{W-RX}\tnote{2}   &
				\textbf{LRASR}\tnote{2}  & 
				\textbf{PAB-DC}\tnote{2} & 
				\textbf{RGAE}\tnote{3}   & 
				\textbf{Auto-AD}\tnote{3} & 
				\textbf{MSBRNet}\tnote{3} & 
				\textbf{S2DWMTrans}\tnote{3} & 
				\textbf{Proposed}\tnote{3} \\
				
				\hline
				
				\multirow{2}{*}{\textbf{Airport I}} & 0.8221 & 0.8876 & 0.8540 & 0.8619 & 0.6388 & \underline{0.9179} & 0.9037 & 0.6616 & \textbf{0.9182} \\
				& 0.12 & 35.00 & 31.80 & 485.79 & 61.85/22.77 & 76.76/13.56 & 329.85/19.83 & 1037.55/48.07 & 92.82/14.49 \\
				\hline
				\multirow{2}{*}{\textbf{Airport II}} & 0.8404 & 0.8412 & 0.8663 & 0.8702 & 0.7478 & 0.8871 & \underline{0.9741} & 0.6965 & \textbf{0.9851} \\
				& 0.11 & 33.11 & 32.73 & 654.64 & 61.65/22.38 & 66.41/8.72 & 314.21/16.13 & 1024.46/44.68 & 92.29/11.71 \\
				\hline
				\multirow{2}{*}{\textbf{Airport III}} & 0.9288 & 0.9367 & 0.9001 & 0.8487 & 0.8877 & 0.8756 & \underline{0.9534} & 0.8465 & \textbf{0.9737} \\
				& 0.11 & 33.55 & 32.50 & 461.73 & 63.75/22.46 & 109.54/11.14 & 314.40/16.26 & 1119.18/46.29 & 92.89/11.58 \\
				\hline
				\multirow{2}{*}{\textbf{Airport IV}} & 0.9526 & 0.7416 & 0.9202 & 0.9056 & 0.7531 & \underline{0.9690} & 0.9540 & 0.9523 & \textbf{0.9966} \\
				& 0.10 & 29.28 & 31.80 & 440.64 & 58.42/20.97 & 53.64/3.73 & 219.22/15.15 & 1013.22/45.29 & 85.57/10.91 \\
				\hline
				\multirow{2}{*}{\textbf{Average}} & 0.8860 & 0.8518 & 0.8851 & 0.8716 & 0.7569 & 0.9124 & \underline{0.9463} & 0.7892 & \textbf{0.9684} \\
				& 0.11 & 32.73 & 32.21 & 510.70 & 61.42/22.14 & 76.59/9.29 & 294.42/16.84 & 1048.60/46.08 & 90.89/12.17 \\
				\hline\hline

				\multirow{2}{*}{\textbf{Beach I}} & \textbf{0.9807} & 0.9702 & 0.9694 & 0.8076 & 0.9392 & 0.9620 & 0.9599 & 0.9728 & \underline{0.9747} \\
				& 0.23 & 68.06 & 92.22 & 2958.41 & 127.72/49.01 & 194.04/14.64 & 680.60/32.08 & 4942.29/139.94 & 185.93/23.76 \\
				\hline
				\multirow{2}{*}{\textbf{Beach II}} & 0.9106 & \textbf{0.9498} & 0.8945 & 0.8915 & 0.9020 & 0.8379 & 0.8739 & 0.9035 & \underline{0.9163} \\
				& 0.10 & 31.05 & 31.87 & 1141.01 & 59.85/21.72 & 86.75/12.29 & 204.93/15.90 & 993.31/46.82 & 85.31/11.10 \\
				\hline
				\multirow{2}{*}{\textbf{Beach III}} & \underline{0.9998} & 0.9997 & 0.9995 & 0.9404 & 0.8667 & 0.9982 & 0.9959 & 0.9884 & \textbf{0.9999} \\
				& 0.09 & 28.39 & 31.77 & 444.08 & 58.77/20.66 & 85.99/9.86 & 195.26/14.76 & 1099.04/44.51 & 84.39/10.82 \\
				\hline
				\multirow{2}{*}{\textbf{Beach IV}} & 0.9538 & 0.9485 & 0.9543 & 0.6462 & 0.9042 & \textbf{0.9900} & 0.9771 & 0.8994 & \underline{0.9871} \\
				& 0.15 & 25.55 & 74.51 & 2095.42 & 92.02/30.88 & 384.31/27.23 & 295.05/18.05 & 5107.37/158.08 & 103.24/13.37 \\
				\hline
				\multirow{2}{*}{\textbf{Average}} & 0.9612 & \underline{0.9670} & 0.9544 & 0.8214 & 0.9030 & 0.9470 & 0.9517 & 0.9410 & \textbf{0.9695} \\
				& 0.14 & 38.26 & 57.59 & 1659.73 & 84.59/30.57 & 187.77/16.00 & 343.96/20.20 & 3035.50/97.34 & 114.72/14.76 \\
				\hline\hline

				\multirow{2}{*}{\textbf{Urban I}} & \underline{0.9907} & 0.9882 & 0.9394 & 0.9685 & 0.9822 & 0.9882 & 0.9460 & 0.9727 & \textbf{0.9915} \\
				& 0.11 & 32.70 & 34.18 & 538.36 & 61.91/22.30 & 77.06/6.44 & 219.45/15.62 & 1085.50/45.31 & 89.99/11.64 \\
				\hline
				\multirow{2}{*}{\textbf{Urban II}} & 0.9946 & 0.9554 & 0.9630 & 0.9799 & \textbf{0.9994} & \underline{0.9992} & 0.8856 & 0.9988 & \textbf{0.9994} \\
				& 0.11 & 35.86 & 28.57 & 478.89 & 61.72/22.64 & 33.12/4.21 & 337.41/16.35 & 1062.00/45.27 & 91.48/11.83 \\
				\hline
				\multirow{2}{*}{\textbf{Urban III}} & 0.9513 & 0.9454 & 0.9392 & 0.9391 & 0.8234 & \textbf{0.9904} & 0.9324 & 0.9238 & \underline{0.9733} \\
				& 0.10 & 31.29 & 32.15 & 507.79 & 58.76/20.88 & 86.69/9.98 & 215.61/15.53 & 997.49/45.15 & 83.88/10.98 \\
				\hline
				\multirow{2}{*}{\textbf{Urban IV}} & 0.9887 & 0.9099 & 0.9529 & 0.7099 & 0.9948 & 0.9903 & 0.9735 & \underline{0.9954} & \textbf{0.9968} \\
				& 0.11 & 35.68 & 20.86 & 443.99 & 61.19/22.19 & 33.12/2.60 & 335.38/16.42 & 990.27/46.45 & 90.42/11.77 \\
				\hline
				\multirow{2}{*}{\textbf{Urban V}} & \underline{0.9692} & 0.9507 & 0.9160 & 0.8161 & 0.9569 & 0.8588 & 0.9162 & 0.9583 & \textbf{0.9757} \\
				& 0.10 & 35.65 & 32.97 & 439.85 & 61.47/22.09 & 120.90/9.99 & 314.20/16.29 & 989.88/44.65 & 91.03/11.75 \\
				\hline
				\multirow{2}{*}{\textbf{Average}} & \underline{0.9789} & 0.9499 & 0.9421 & 0.8827 & 0.9513 & 0.9654 & 0.9307 & 0.9698 & \textbf{0.9873} \\
				& 0.11 & 34.24 & 29.75 & 481.78 & 61.01/22.02 & 70.18/6.64 & 284.41/16.04 & 1025.03/45.37 & 89.36/11.59 \\
				\hline\hline

				\multirow{2}{*}{ \makecell{ \textbf{HyMap}\\\textbf{Cooke City}\tnote{4} } } & 0.8171 & 0.6192 & \underline{0.8864} & - & 0.7603 & 0.5312 & 0.5614 & - & \textbf{0.9067} \\
				& 1.47 & 338.72 & 1621.33 & - & 1091.07/676.17 & 2522.17/118.33 & 4472.95/190.81 & -/- & 1350.14/172.32 \\
				
				\bottomrule
			\end{tabular}
			
			\begin{tablenotes}
				\footnotesize
				\item[1] The best AUC score among all methods on one dataset or average is in \textbf{bold}, and the second is \underline{underlined}.
				\item[2] The computational time of these non-deep learning-based methods are obtained by running the algorithms on the CPU.
				\item[3] These deep learning-based methods are run on both CPU and GPU devices. We list the computational time as CPU/GPU.
				\item[4] Our hardware devices do not have sufficient capacity to run the PAB-DC and S2DWMTrans algorithms because the size of this dataset is too large.
			\end{tablenotes}
			
		\end{threeparttable}
	\end{table*}
	
	To quantitatively compare the performance, we list the AUC scores obtained by all considered methods on ABU and HyMap Cooke City datasets in Table \ref{tab_performance_comparison}. The best and second-best results for each dataset are highlighted in \textbf{bold} and \underline{underlined}, respectively. We also include the computational time required by each method in the table to provide an evaluation of its complexity. We run all algorithms on the CPU, and for deep learning-based methods, we also perform an additional test on the GPU. 
	
	As shown in Table \ref{tab_performance_comparison}, our proposed method outperforms all 8 competitive methods on 10 datasets, including all 4 Airport datasets and the Urban I, II, IV, and V datasets, as well as Beach III and HyMap Cooke City. The RX, W-RX, and Auto-AD methods achieve the highest scores on Beach I, II, and IV and Urban III, while our method ranks second. Overall, our method achieves the best average performance across all airport, beach, and urban datasets. Besides, it surpasses all competitive methods on HyMap Cooke City, which has the largest size and the most complex scene. These results demonstrate the effectiveness of our method for anomaly detection and its adaptability to various scenes.
	
	Among all methods run on the CPU, the RX algorithm has the lowest computational time on all datasets, while the S2DWMTrans method is the most time-consuming due to its complex network. For deep learning-based methods run on the GPU, the Auto-AD algorithm is the fastest, and our method ranks second on average.
	
	\begin{figure*}[!t]
		\centering
		\includegraphics[scale=0.55]{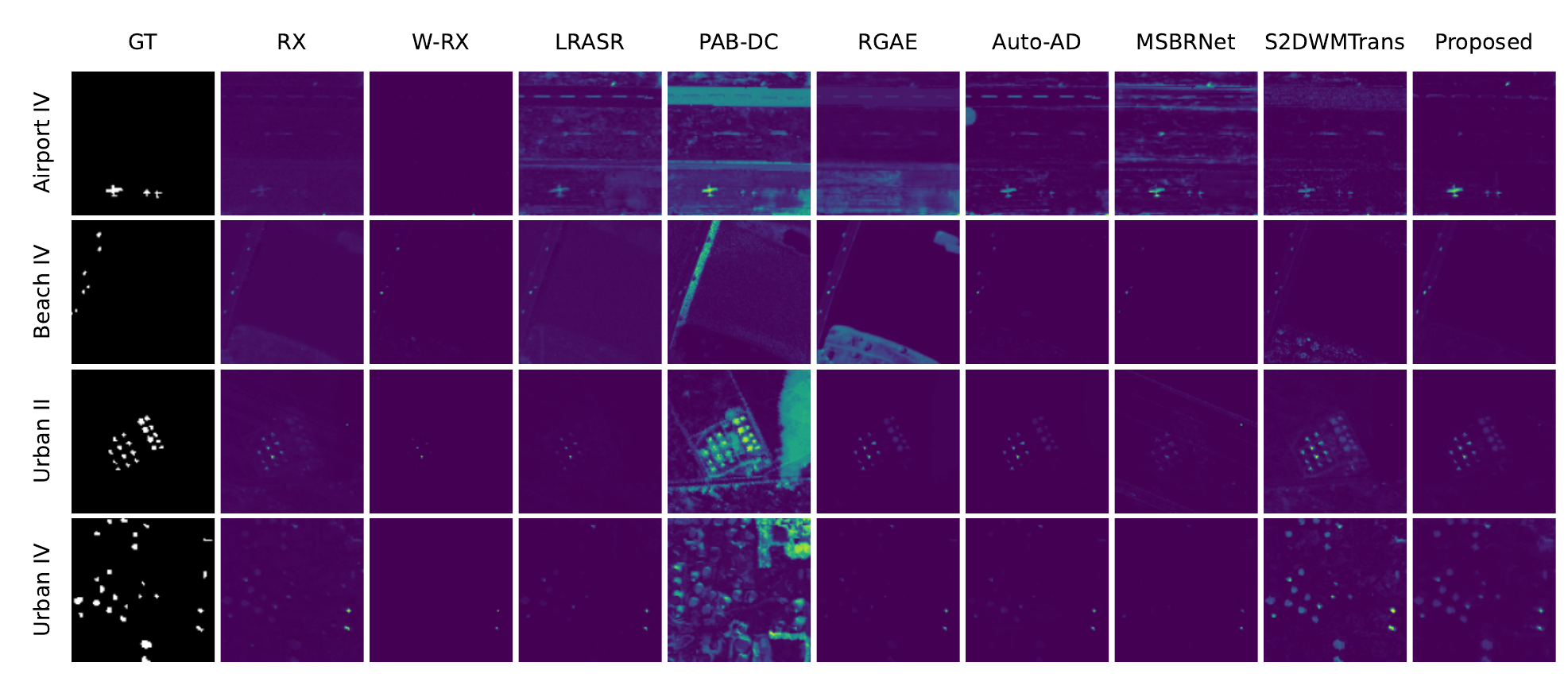}
		\caption{Ground truth maps (first column) and corresponding detection maps for different methods (remaining columns) on 4 analyzed datasets. (a) Airport IV. (b) Beach IV. (c) Urban II. (d) Urban IV. }
		\label{fig_abu_pc}
	\end{figure*}
	
	For visual comparisons, the detection maps generated by 9 methods on 4 analyzed datasets Airport IV, Beach IV, Urban II and Urban IV, are shown in Fig. \ref{fig_abu_pc}. It can be seen that our method demonstrates the clearest detection results among all 9 methods, with the background being well removed and most anomalies being highlighted. Taking the Airport IV dataset as an example, some methods including RX, W-RX and RGAE, cannot clearly mark all three airplanes in this scene, while some methods including PAB-DC, MSBRNet, and S2DWMTrans mix obvious background components in their detection maps. Contrarily, our method can completely detect the objects of interest, and keep remaining areas cleaner.
	
	\begin{figure}[!t]
		\centering
		\subfloat[]{\includegraphics[scale=0.28]{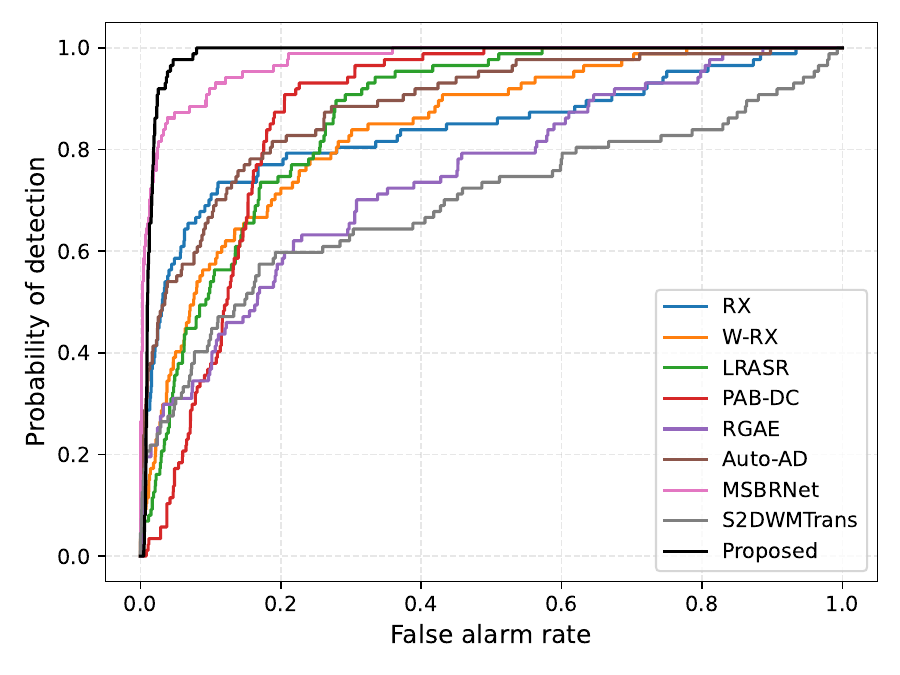}%
			\label{fig_a2_roc}}
		\hfil
		\subfloat[]{\includegraphics[scale=0.28]{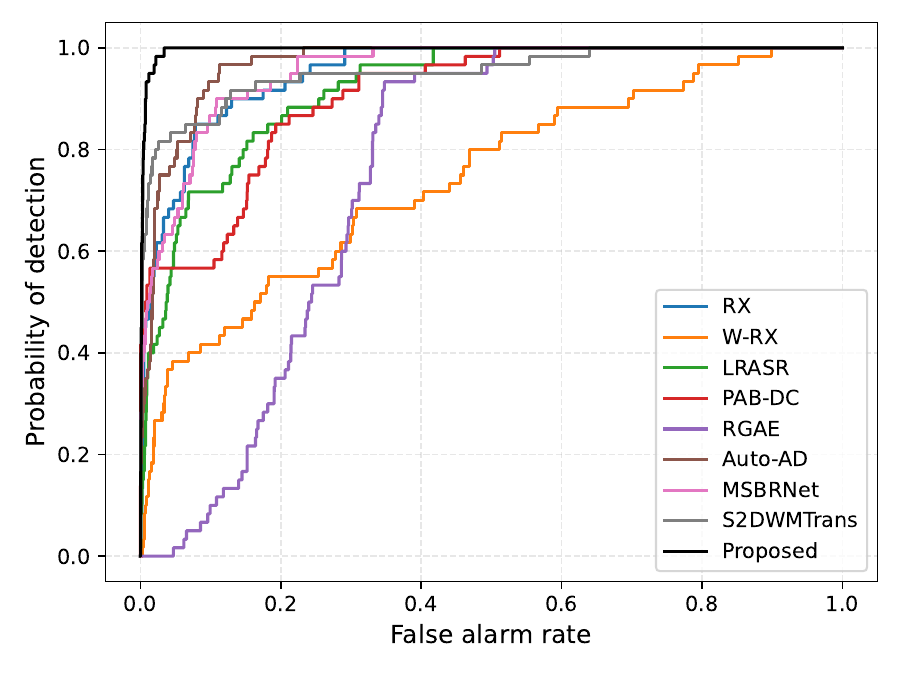}%
			\label{fig_a4_roc}}
		\hfil
		\subfloat[]{\includegraphics[scale=0.28]{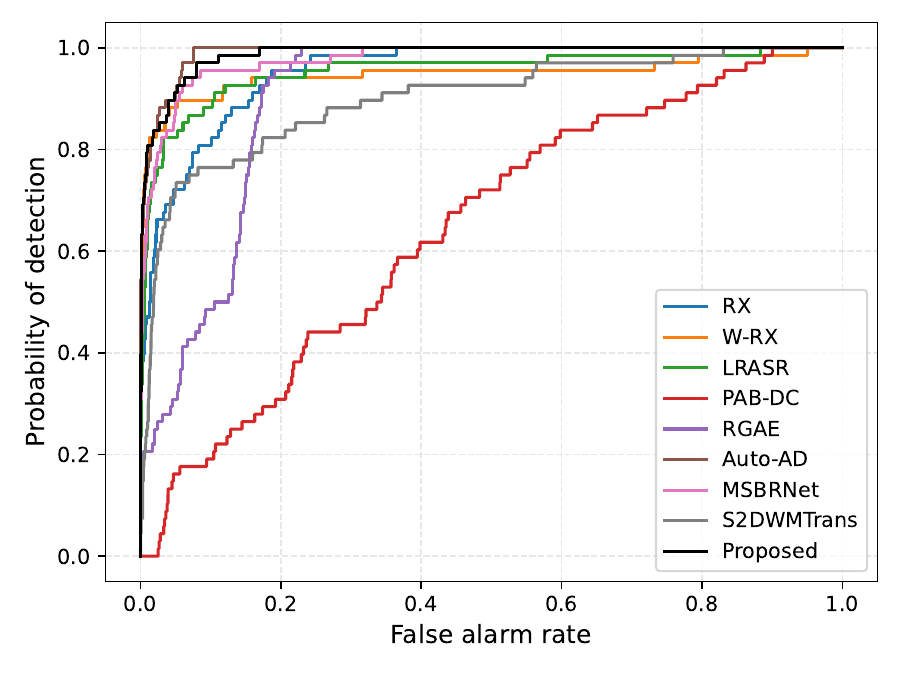}%
			\label{fig_b4_roc}}
		\hfil
		\subfloat[]{\includegraphics[scale=0.28]{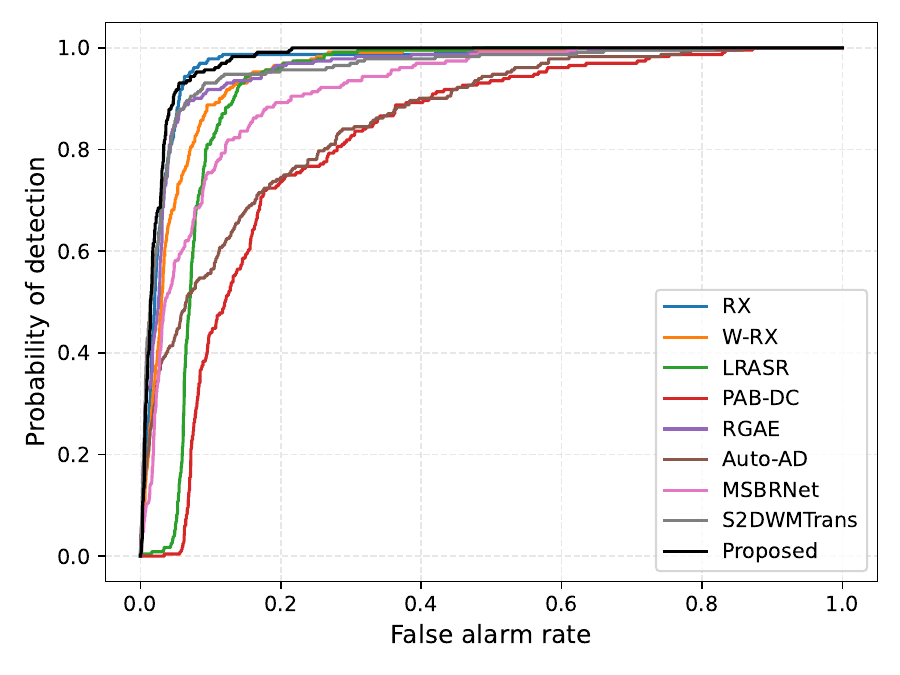}%
			\label{fig_u5_roc}}
		\caption{ROC curves for different algorithms on 4 analyzed datasets. (a) Airport II. (b) Airport IV. (c) Beach IV. (d) Urban V.}
		\label{fig_roc}
	\end{figure}
	
	The ROC curves obtained by all methods on 4 selected datasets, i.e. Airport II, Airport IV, Beach IV and Urban V, are respectively shown in Fig. \ref{fig_roc} (a-d). It can be observed in Fig. \ref{fig_roc} (a) and (b) that the ROC curves of our method (in black) are closer to the top-left corner compared to other methods, indicating higher detection performance. In Fig. \ref{fig_roc} (c), the ROC curve of the Auto-AD displays a slightly higher probability of detection than our method when the false alarm rate is between 0 and close to 0.2. In Fig. \ref{fig_roc} (b), our method and the RX show similar trends and almost cover the ROC curves of other methods.

	\subsection{Transferability Discussion}
	
	The separation training strategy is proposed to address the IMP of deep networks, as discussed in Section III. Because our \proposed~is a model-independent training strategy (you can replace the AE structure with an arbitrary model whose input and output have the same size, and the training will still work), we validate its effectiveness by comparing the improvement of a network trained with this strategy to the original one. The validation simultaneously exhibits the transferability of the strategy from a simple AE to deeper networks. 
	
	\subsubsection{Experimental Setup}
	
	The candidate networks are two CNNs Auto-AD and MSBRNet, the ViT S2DWMTrans and the vanilla AE. In addition to our proposed separation training strategy (denoted as BiGSeT) for anomaly suppression, we also compare two other strategies, i.e. the $l_{2, 1}$-norm \cite{ref_rgae} (denoted as $l_{2, 1}$) and the adaptive-weighted training \cite{ref_auto_ad} (denoted as AW), respectively. The training settings for the Auto-AD, MSBRNet and S2DWMTrans are followed according to their original papers, and we train the original AE with the Frobenius norm loss for 750 epochs. 
	
	In the following experiments, we will combine the 4 networks with the 4 training methods, resulting in a total of 16 combinations (actually 15 because the original Auto-AD adopts the AW strategy), to compare the performance of the original version with the other 3 training strategies. For a fair comparison, we terminate the training processes at the 750th epoch for all networks except the original versions, and obtain the detection results accordingly.
	
	\subsubsection{Quantitative Results}
	
	\begin{table*}[!t]
		\caption{
			AUC Scores (Upper Row) and Improvement on Networks by Anomaly Suppression (Lower Row, Compared with Their Original Results) on All Datasets
			\label{tab_boost_by_anomaly_suppression}
		}
		\renewcommand{\arraystretch}{1.5}
		\scriptsize
		\tabcolsep=0.10cm
		\centering
		\begin{threeparttable}
			\begin{tabular}{c | c c c c | c c c c | c c c c | c c c c}
				\toprule
				
				\multirow{2}{*}{\textbf{Dataset}} & \multicolumn{4}{c|}{\textbf{Auto-AD}} & \multicolumn{4}{c|}{\textbf{MSBRNet}} & \multicolumn{4}{c|}{\textbf{S2DWMTrans}} & \multicolumn{4}{c}{\textbf{AE}} \\
				\cline{2-17}
				& Orig.\tnote{1} & $l_{2, 1}$\tnote{1} & AW\tnote{1} & BiGSeT\tnote{1} & Orig. & $l_{2, 1}$ & AW & BiGSeT & Orig. & $l_{2, 1}$ & AW & BiGSeT & Orig. & $l_{2, 1}$ & AW & BiGSeT \\
				
				\hline
				
				\multirow{2}{*}{\textbf{Airport I}} & \cellcolor{lightgray}0.9179 & 0.8392 & 0.9179 & \textbf{0.9192} & \cellcolor{lightgray}0.9037 & 0.8893 & 0.8788 & \textbf{0.9069} & \cellcolor{lightgray}0.6616 & 0.7997 & 0.8809 & \textbf{0.9315} & \cellcolor{lightgray}0.8808 & 0.8740 & 0.8902 & \textbf{0.9182} \\
				& \cellcolor{lightgray}- & -0.0787 & - & \textbf{+0.0013} & \cellcolor{lightgray}- & -0.0144 & -0.0249 & \textbf{+0.0032} & \cellcolor{lightgray}- & +0.1381 & +0.2193 & \textbf{+0.2699} & \cellcolor{lightgray}- & -0.0068 & +0.0094 & \textbf{+0.0374} \\
				\hline
				
				\multirow{2}{*}{\textbf{Airport II}} & \cellcolor{lightgray}0.8871 & 0.8667 & 0.8871 & \textbf{0.9694} & \cellcolor{lightgray}0.9741 & 0.9697 & 0.9517 & \textbf{0.9783} & \cellcolor{lightgray}0.6965 & 0.9056 & 0.9301 & \textbf{0.9873} & \cellcolor{lightgray}0.9324 & 0.9229 & 0.9312 & \textbf{0.9851} \\
				& \cellcolor{lightgray}- & -0.0204 & - & \textbf{+0.0823} & \cellcolor{lightgray}- & -0.0044 & -0.0224 & \textbf{+0.0042} & \cellcolor{lightgray}- & +0.2091 & +0.2336 & \textbf{+0.2908} & \cellcolor{lightgray}- & -0.0095 & -0.0012 & \textbf{+0.0527} \\
				\hline
				
				\multirow{2}{*}{\textbf{Airport III}} & \cellcolor{lightgray}0.8756 & 0.8972 & 0.8756 & \textbf{0.9132} & \cellcolor{lightgray}0.9534 & 0.9077 & 0.9198 & \textbf{0.9659} & \cellcolor{lightgray}0.8465 & 0.9035 & 0.9269 & \textbf{0.9579} & \cellcolor{lightgray}0.9190 & 0.9181 & 0.9245 & \textbf{0.9737} \\
				& \cellcolor{lightgray}- & +0.0216 & - & \textbf{+0.0376} & \cellcolor{lightgray}- & -0.0457 & -0.0336 & \textbf{+0.0125} & \cellcolor{lightgray}- & +0.0570 & +0.0804 & \textbf{+0.1114} & \cellcolor{lightgray}- & -0.0009 & +0.0055 & \textbf{+0.0547} \\
				\hline
				
				\multirow{2}{*}{\textbf{Airport IV}} & \cellcolor{lightgray}0.9690 & 0.9534 & 0.9690 & \textbf{0.9936} & \cellcolor{lightgray}0.9540 & 0.9935 & 0.9649 & \textbf{0.9984} & \cellcolor{lightgray}0.9523 & 0.9760 & 0.9883 & \textbf{0.9991} & \cellcolor{lightgray}0.9889 & 0.9899 & 0.9918 & \textbf{0.9966} \\
				& \cellcolor{lightgray}- & -0.0156 & - & \textbf{+0.0246} & \cellcolor{lightgray}- & +0.0395 & +0.0109 & \textbf{+0.0444} & \cellcolor{lightgray}- & +0.0237 & +0.0360 & \textbf{+0.0468} & \cellcolor{lightgray}- & +0.0010 & +0.0029 & \textbf{+0.0077} \\
				\hline
				
				\multirow{2}{*}{\textbf{Average}} & \cellcolor{lightgray}0.9124 & 0.8891 & 0.9124 & \textbf{0.9488} & \cellcolor{lightgray}0.9463 & 0.9401 & 0.9288 & \textbf{0.9624} & \cellcolor{lightgray}0.7892 & 0.8962 & 0.9315 & \textbf{0.9689} & \cellcolor{lightgray}0.9303 & 0.9262 & 0.9344 & \textbf{0.9684} \\
				& \cellcolor{lightgray}- & -0.0233 & - & \textbf{+0.0364} & \cellcolor{lightgray}- & -0.0062 & -0.0175 & \textbf{+0.0161} & \cellcolor{lightgray}- & +0.1070 & +0.1423 & \textbf{+0.1797} & \cellcolor{lightgray}- & -0.0041 & +0.0041 & \textbf{+0.0381} \\
				\hline\hline
				
				\multirow{2}{*}{\textbf{Beach I}} & \cellcolor{lightgray}0.9620 & 0.9910 & 0.9620 & \textbf{0.9975} & \cellcolor{lightgray}0.9599 & 0.9792 & 0.9822 & \textbf{0.9860} & \cellcolor{lightgray}0.9728 & \textbf{0.9910} & 0.9812 & 0.9848 & \cellcolor{lightgray}0.9710 & \textbf{0.9875} & 0.9825 & 0.9747 \\
				& \cellcolor{lightgray}- & +0.0290 & - & \textbf{+0.0355} & \cellcolor{lightgray}- & +0.0193 & +0.0223 & \textbf{+0.0261} & \cellcolor{lightgray}- & \textbf{+0.0182} & +0.0084 & +0.0120 & \cellcolor{lightgray}- & \textbf{+0.0165} & +0.0115 & +0.0037 \\
				\hline
				
				\multirow{2}{*}{\textbf{Beach II}} & \cellcolor{lightgray}0.8379 & 0.8982 & 0.8379 & \textbf{0.9007} & \cellcolor{lightgray}0.8739 & \textbf{0.9027} & 0.8734 & 0.8859 & \cellcolor{lightgray}0.9035 & 0.8908 & 0.8899 & \textbf{0.9046} & \cellcolor{lightgray}0.9159 & 0.9150 & 0.9138 & \textbf{0.9163} \\
				& \cellcolor{lightgray}- & +0.0603 & - & \textbf{+0.0628} & \cellcolor{lightgray}- & \textbf{+0.0288} & -0.0005 & +0.0120 & \cellcolor{lightgray}- & -0.0127 & -0.0136 & \textbf{+0.0011} & \cellcolor{lightgray}- & -0.0009 & -0.0021 & \textbf{+0.0004} \\
				\hline
				
				\multirow{2}{*}{\textbf{Beach III}} & \cellcolor{lightgray}0.9982 & 0.9882 & 0.9982 & \textbf{0.9999} & \cellcolor{lightgray}0.9959 & 0.9927 & 0.9925 & \textbf{0.9995} & \cellcolor{lightgray}0.9884 & 0.9998 & \textbf{1.0000} & 0.9999 & \cellcolor{lightgray}0.9998 & \textbf{0.9999} & \textbf{0.9999} & \textbf{0.9999} \\
				& \cellcolor{lightgray}- & -0.0100 & - & \textbf{+0.0017} & \cellcolor{lightgray}- & -0.0032 & -0.0034 & \textbf{+0.0036} & \cellcolor{lightgray}- & +0.0114 & \textbf{+0.0116} & +0.0115 & \cellcolor{lightgray}- & \textbf{+0.0001} & \textbf{+0.0001} & \textbf{+0.0001} \\
				\hline
				
				\multirow{2}{*}{\textbf{Beach IV}} & \cellcolor{lightgray}0.9900 & 0.9765 & 0.9900 & \textbf{0.9931} & \cellcolor{lightgray}0.9771 & 0.9565 & 0.9761 & \textbf{0.9919} & \cellcolor{lightgray}0.8994 & 0.9636 & 0.9595 & \textbf{0.9946} & \cellcolor{lightgray}0.9663 & 0.9516 & 0.9709 & \textbf{0.9871} \\
				& \cellcolor{lightgray}- & -0.0135 & - & \textbf{+0.0031} & \cellcolor{lightgray}- & -0.0206 & -0.0010 & \textbf{+0.0148} & \cellcolor{lightgray}- & +0.0642 & +0.0601 & \textbf{+0.0952} & \cellcolor{lightgray}- & -0.0147 & +0.0046 & \textbf{+0.0208} \\
				\hline
				
				\multirow{2}{*}{\textbf{Average}} & \cellcolor{lightgray}0.9470 & 0.9635 & 0.9470 & \textbf{0.9728} & \cellcolor{lightgray}0.9517 & 0.9578 & 0.9561 & \textbf{0.9658} & \cellcolor{lightgray}0.9410 & 0.9613 & 0.9577 & \textbf{0.9710} & \cellcolor{lightgray}0.9633 & 0.9635 & 0.9668 & \textbf{0.9695} \\
				& \cellcolor{lightgray}- & +0.0164 & - & \textbf{+0.0258} & \cellcolor{lightgray}- & +0.0061 & +0.0044 & \textbf{+0.0141} & \cellcolor{lightgray}- & +0.0203 & +0.0166 & \textbf{+0.0300} & \cellcolor{lightgray}- & +0.0002 & +0.0035 & \textbf{+0.0062} \\
				\hline\hline
				
				\multirow{2}{*}{\textbf{Urban I}} & \cellcolor{lightgray}0.9882 & 0.9587 & 0.9882 & \textbf{0.9978} & \cellcolor{lightgray}0.9460 & 0.9257 & 0.9727 & \textbf{0.9844} & \cellcolor{lightgray}0.9727 & 0.9824 & 0.9818 & \textbf{0.9929} & \cellcolor{lightgray}0.9701 & 0.9881 & 0.9809 & \textbf{0.9915} \\
				& \cellcolor{lightgray}- & -0.0295 & - & \textbf{+0.0096} & \cellcolor{lightgray}- & -0.0203 & +0.0267 & \textbf{+0.0384} & \cellcolor{lightgray}- & +0.0097 & +0.0091 & \textbf{+0.0202} & \cellcolor{lightgray}- & +0.0180 & +0.0108 & \textbf{+0.0214} \\
				\hline
				
				\multirow{2}{*}{\textbf{Urban II}} & \cellcolor{lightgray}\textbf{0.9992} & 0.9008 & \textbf{0.9992} & \textbf{0.9992} & \cellcolor{lightgray}0.8856 & 0.9950 & 0.9160 & \textbf{0.9974} & \cellcolor{lightgray}0.9988 & 0.9942 & 0.9842 & \textbf{0.9990} & \cellcolor{lightgray}0.9199 & 0.9829 & 0.9043 & \textbf{0.9994} \\
				& \cellcolor{lightgray}- & -0.0984 & - & \textbf{+0.0000} & \cellcolor{lightgray}- & +0.1094 & +0.0304 & \textbf{+0.1118} & \cellcolor{lightgray}- & -0.0046 & -0.0146 & \textbf{+0.0002} & \cellcolor{lightgray}- & +0.0630 & -0.0156 & \textbf{+0.0795} \\
				\hline
				
				\multirow{2}{*}{\textbf{Urban III}} & \cellcolor{lightgray}0.9904 & 0.9513 & 0.9904 & \textbf{0.9925} & \cellcolor{lightgray}0.9324 & \textbf{0.9794} & 0.9344 & 0.9725 & \cellcolor{lightgray}0.9238 & 0.9377 & 0.9622 & \textbf{0.9710} & \cellcolor{lightgray}0.9553 & 0.9615 & 0.9613 & \textbf{0.9733} \\
				& \cellcolor{lightgray}- & -0.0391 & - & \textbf{+0.0021} & \cellcolor{lightgray}- & \textbf{+0.0470} & +0.0020 & +0.0401 & \cellcolor{lightgray}- & +0.0139 & +0.0384 & \textbf{+0.0472} & \cellcolor{lightgray}- & +0.0062 & +0.0060 & \textbf{+0.0180} \\
				\hline
				
				\multirow{2}{*}{\textbf{Urban IV}} & \cellcolor{lightgray}0.9903 & 0.9346 & 0.9903 & \textbf{0.9970} & \cellcolor{lightgray}0.9735 & 0.9434 & 0.9650 & \textbf{0.9967} & \cellcolor{lightgray}0.9954 & 0.9893 & 0.9726 & \textbf{0.9970} & \cellcolor{lightgray}0.9562 & 0.9721 & 0.9653 & \textbf{0.9968} \\
				& \cellcolor{lightgray}- & -0.0557 & - & \textbf{+0.0067} & \cellcolor{lightgray}- & -0.0301 & -0.0085 & \textbf{+0.0232} & \cellcolor{lightgray}- & -0.0061 & -0.0228 & \textbf{+0.0016} & \cellcolor{lightgray}- & +0.0159 & +0.0091 & \textbf{+0.0406} \\
				\hline
				
				\multirow{2}{*}{\textbf{Urban V}} & \cellcolor{lightgray}0.8588 & 0.8961 & 0.8588 & \textbf{0.9726} & \cellcolor{lightgray}0.9162 & 0.9221 & 0.9001 & \textbf{0.9600} & \cellcolor{lightgray}0.9583 & 0.9639 & 0.9666 & \textbf{0.9760} & \cellcolor{lightgray}0.9544 & 0.9561 & 0.9651 & \textbf{0.9757} \\
				& \cellcolor{lightgray}- & +0.0373 & - & \textbf{+0.1138} & \cellcolor{lightgray}- & +0.0059 & -0.0161 & \textbf{+0.0438} & \cellcolor{lightgray}- & +0.0056 & +0.0083 & \textbf{+0.0177} & \cellcolor{lightgray}- & +0.0017 & +0.0107 & \textbf{+0.0213} \\
				\hline
				
				\multirow{2}{*}{\textbf{Average}} & \cellcolor{lightgray}0.9654 & 0.9283 & 0.9654 & \textbf{0.9918} & \cellcolor{lightgray}0.9307 & 0.9531 & 0.9376 & \textbf{0.9822} & \cellcolor{lightgray}0.9698 & 0.9735 & 0.9735 & \textbf{0.9872} & \cellcolor{lightgray}0.9512 & 0.9721 & 0.9554 & \textbf{0.9873} \\
				& \cellcolor{lightgray}- & -0.0371 & - & \textbf{+0.0264} & \cellcolor{lightgray}- & +0.0224 & +0.0069 & \textbf{+0.0515} & \cellcolor{lightgray}- & +0.0037 & +0.0037 & \textbf{+0.0174} & \cellcolor{lightgray}- & +0.0210 & +0.0042 & \textbf{+0.0362} \\
				\hline\hline
				
				\multirow{2}{*}{\makecell{\textbf{HyMap}\\\textbf{Cooke City}\tnote{2}}} & \cellcolor{lightgray}0.5312 & 0.5250 & 0.4609 & \textbf{0.6028} & \cellcolor{lightgray}0.5614 & 0.6896 & 0.6586 & \textbf{0.7846} & \cellcolor{lightgray}- & - & - & - & \cellcolor{lightgray}0.8499 & 0.7900 & 0.8945 & \textbf{0.9067} \\
				& \cellcolor{lightgray}- & -0.0062 & -0.0703 & \textbf{+0.0716} & \cellcolor{lightgray}- & +0.1282 & +0.0972 & \textbf{+0.2232} & \cellcolor{lightgray}- & - & - & - & \cellcolor{lightgray}- & -0.0599 & +0.0446 & \textbf{+0.0568} \\
				
				\bottomrule
			\end{tabular}
			
			\begin{tablenotes}
				\footnotesize
				\item[1] We compare the AUC improvement of three model-independent anomaly suppression strategies on the original networks to verify the effectiveness and transferability of our method. The highest AUC score and largest improvement of one network on one dataset are in \textbf{bold}.
				\item[2] The results are unavailable on this dataset for the S2DWMTrans network due to the large size of the image.
			\end{tablenotes}
			
		\end{threeparttable}
	\end{table*}
	
	The detection results of all combinations on ABU datasets and HyMap Cooke City dataset are listed in Table \ref{tab_boost_by_anomaly_suppression}. For a clearer comparison, we also provide the improvement scores of the other three training strategies over the original ones. We can observe that our BiGSeT strategy leads to positive improvements in all cases and achieves the largest improvement on most datasets. However, the $l_{2, 1}$ or AW strategies display worse results than the original on some datasets, such as Airport II for the Auto-AD and MSBRNet networks. With the improvement of the BiGSeT, detection results can even be raised to a very high degree. For instance, the combination of S2DWMTrans with BiGSeT can raise the AUC score on Airport IV from 0.9523 to 0.9991. To conclude, our proposed BiGSeT strategy can improve the detection performance of 4 tested networks and outperforms the other two anomaly suppression methods, $l_{2, 1}$ and AW.
	
	\subsubsection{Training Process Curves}
	
	\begin{figure*}[!t]
		\centering
		\includegraphics[scale=0.47]{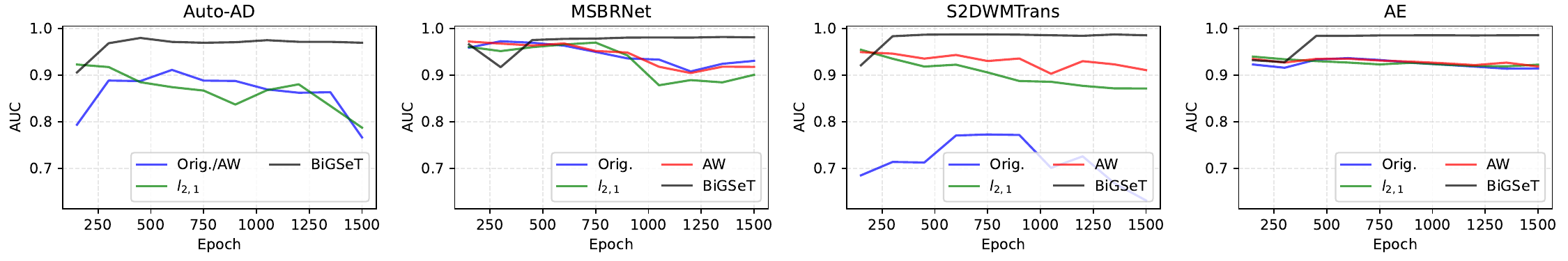}
		\caption{Training processes using different strategies (AUC curves in blue, red, green and black) for 4 networks on Airport II dataset.}
		\label{fig_a2_tp}
	\end{figure*}
	
	\begin{figure*}[!t]
		\centering
		\includegraphics[scale=0.47]{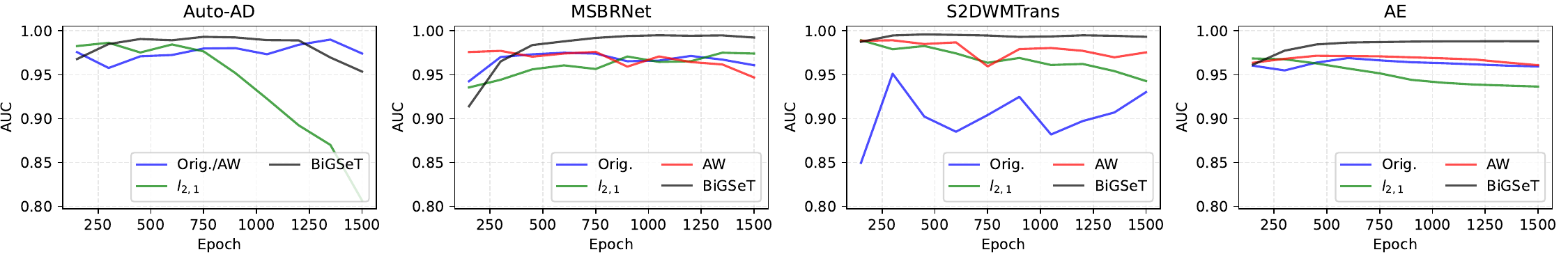}
		\caption{Training processes using different strategies (AUC curves in blue, red, green and black) for 4 networks on Beach IV dataset.}
		\label{fig_b4_tp}
	\end{figure*}
	
	\begin{figure*}[!t]
		\centering
		\includegraphics[scale=0.47]{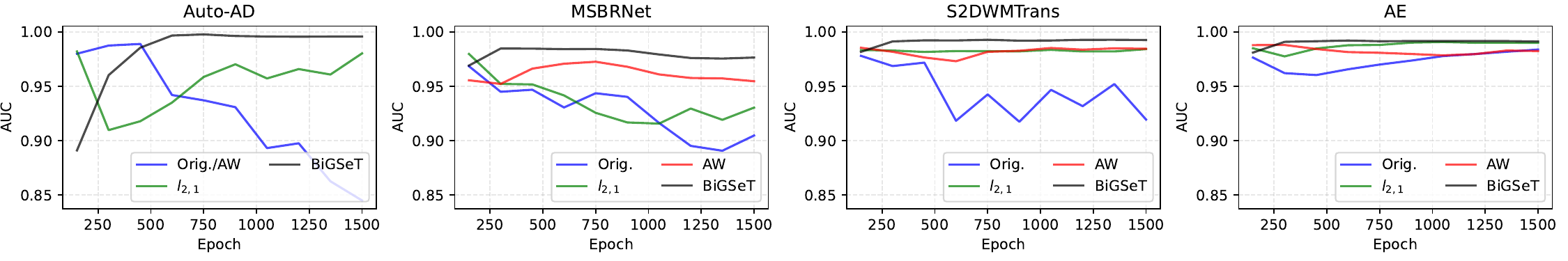}
		\caption{Training processes using different strategies (AUC curves in blue, red, green and black) for 4 networks on Urban I dataset.}
		\label{fig_u1_tp}
	\end{figure*}
	
	To better explain how these training strategies impact on detection performance, we display the process curves of the 4 networks along the training dimension on three datasets, as shown in Figs. \ref{fig_a2_tp}-\ref{fig_u1_tp}. It can be observed from the 3 figures that the AUC scores of the most original networks (in blue) tend to decline or fluctuate as training progresses. This is because the networks start to fit the image to a high degree, and the anomalies and background fail to be distinguished. This phenomenon seems more significant to the Auto-AD, MSBRNet and S2DWMTrans because their structures are more complex to easily overfit learning. To prevent this, the training strategies try to lift and flatten the curves (in green for $l_{2, 1}$, red for AW except blue for the Auto-AD, and black for ours), by suppressing the anomalies from being fitted. Although the $l_{2, 1}$ and AW methods show improvement in certain cases, such as using S2DWMTrans on the Airport II dataset, they fail to maintain high performance as training progresses. The $l_{2, 1}$ method lacks the ability to provide positional information of the anomalous targets, resulting in ineffective suppression of these pixels. On the other hand, the AW method employs a soft weight map that may lead to blurry separation and does not ensure a balanced equilibrium for maintaining separability. In contrast, our method separately constrains the background and anomaly parts using a binary mask, explicitly removing anomalies to the greatest extent possible. Additionally, the mask is robustly updated using a proportion threshold, which guarantees that the identified anomalies will not be learned during training. As a result, the performance does not decrease in most cases, and our proposed BiGSeT strategy provides faster-converged and stabler results among all these methods.
	
 	\begin{table}[!t]
		\caption{
			Network Parameters (In Millions) / FLOPs (In Billions) for Four Networks on Three Datasets
			\label{tab_params_flops_comparison}
		}
		\renewcommand{\arraystretch}{1.5}
		\tabcolsep=0.11cm
		\centering
		\begin{tabular}{c c c c c}
			\toprule
			\textbf{Dataset}    & \textbf{Auto-AD}    & \textbf{MSBRNet} & \textbf{S2DWMTrans}  & \textbf{AE}  \\
			\hline
			\textbf{Airport IV} & 3.23 / \enspace5.92 & 1.83 / 18.76     & 0.51 / \enspace56.29 & 0.04 / 0.38 \\
			\textbf{Beach I}    & 3.23 / 13.28        & 1.78 / 41.08     & 0.51 / 270.57        & 0.04 / 0.85  \\
			\textbf{Urban IV}   & 3.25 / \enspace5.99 & 2.07 / 21.20     & 0.53 / \enspace56.47 & 0.04 / 0.41 \\
			
			\bottomrule
		\end{tabular}
	\end{table}
 
    \subsubsection{Complexity of Networks}	
	We tabulate the parameters and FLOPs of 4 networks on 3 datasets in Table \ref{tab_params_flops_comparison} in order to compare their depth and complexity. According to Table \ref{tab_params_flops_comparison}, the Auto-AD has the largest parameter size and the S2DWMTrans is the network with the highest computational cost, while the AE is the simplest structure.
	
	\subsection{Hyperparameter Analysis}
	
	In this subsection, we explore how the hyperparameters $\gamma$ and $\lambda$ affect the model. As an example, we conduct parameter analysis on the Airport IV dataset. All the experiments are conducted with a vanilla AE structure introduced before. 
	
	\begin{figure}[!t]
		\centering
		\subfloat[]{\includegraphics[scale=0.24]{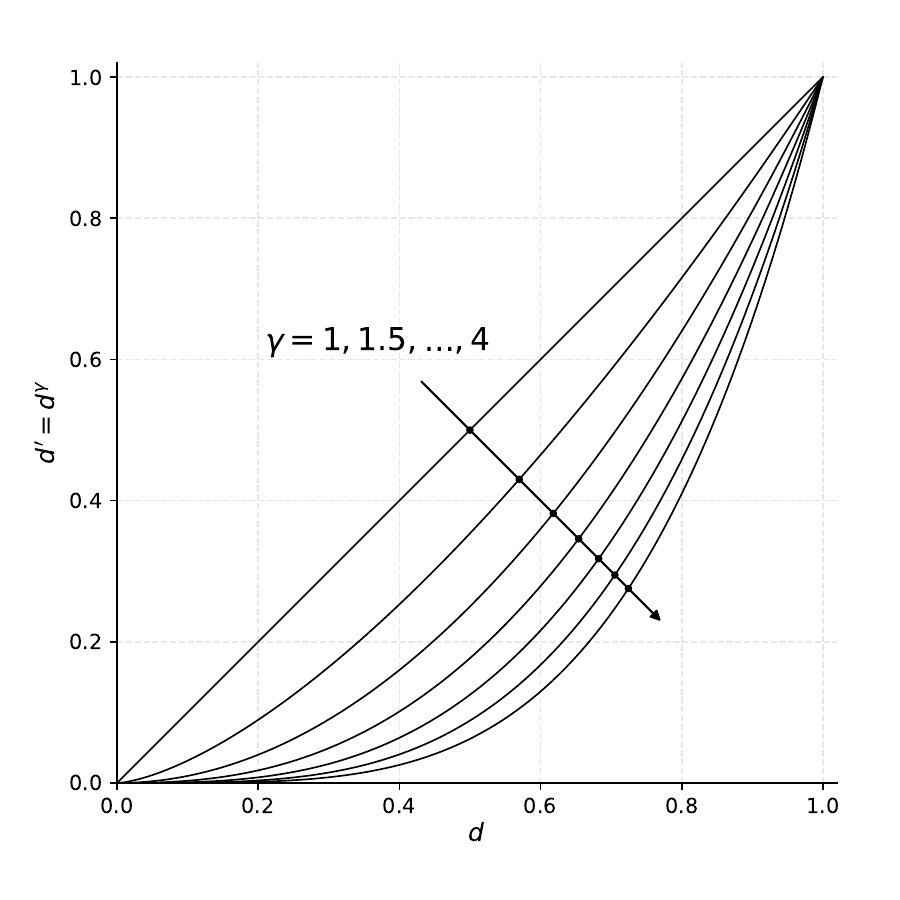}%
			\label{fig_gamma_curve}}
		\hfil
		\subfloat[]{\includegraphics[scale=0.24]{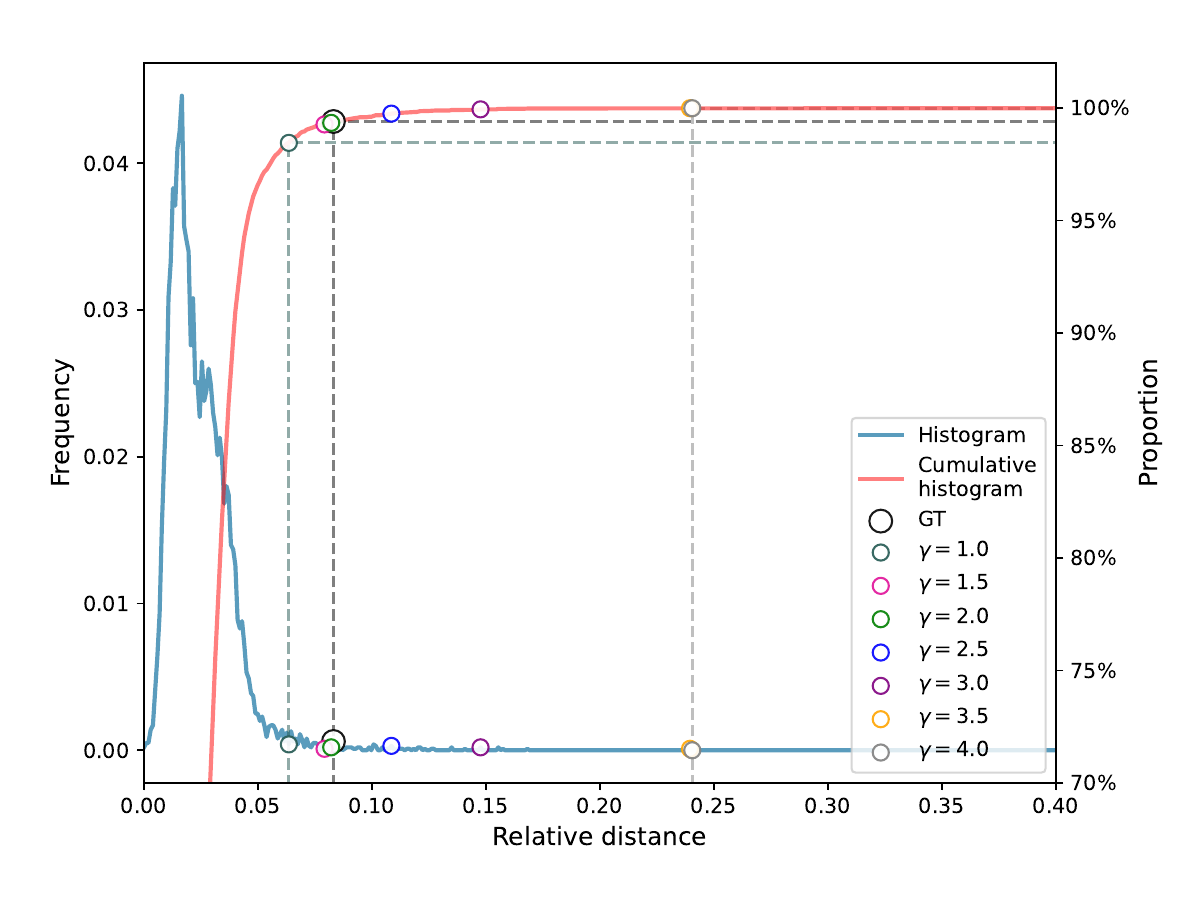}%
			\label{fig_A4_hist_thresh}}
		\caption{Impact of $\gamma$ on threshold estimation. (a) Gamma curves for different values. (b) Histogram and cumulative histogram curves of relative distance on Airport IV dataset. We use different colors to mark the estimated thresholds with respect to $\gamma$, along with the ground truth.}
		\label{fig_thresh_est}
	\end{figure}

    \begin{figure}[!t]
		\centering
		\includegraphics[scale=0.5]{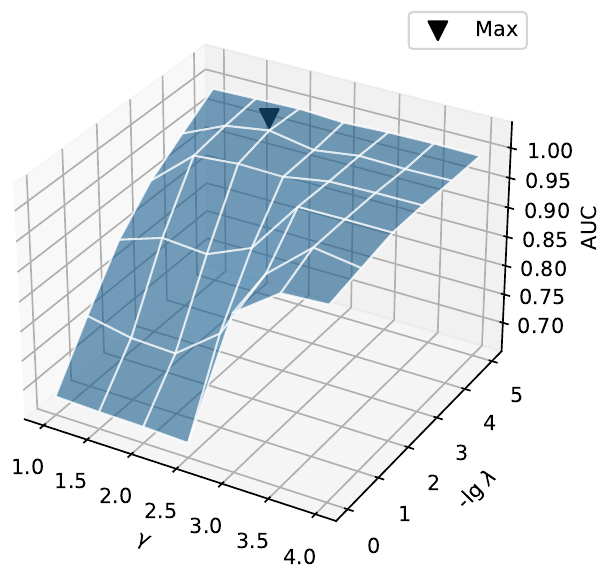}
		\caption{Joint impact of $\gamma$ and $\lambda$ on Airport IV dataset.}
		\label{fig_a4_params}
	\end{figure}
	
	The parameter $\gamma$ controls the degree of convexity of gamma transformation curves, as shown in Fig. \ref{fig_thresh_est} (a). The larger $\gamma$ is, the more intensive the effect of mapping $d$ values to smaller and narrower ranges is. We test different $\gamma$ values chosen from $\{1.0, 1.5, 2.0, 2.5, 3.0, 3.5, 4.0\}$ to display its impact on the proportion threshold estimation, as shown in Fig. \ref{fig_thresh_est} (b). The histogram of relative distance exhibits a unimodal characteristic, while its corresponding cumulative histogram initially shows a steep rise and then levels off, as the proportion of anomalous targets is relatively low. The ground truth proportion threshold is located near the corner. By adjusting the gamma transformation, the proportion threshold estimation is closer to the ground truth when the $\gamma$ values are set to 1.5 and 2.0.
	
    The parameter $\lambda$ balances the effects of background reconstruction and anomaly suppression. To get the best result and better understand the two parameters, we investigate the joint impact of $\gamma$ and $\lambda$ on the AUC score. We choose their values respectively from candidate pools $\{1.0, 1.5, 2.0, 2.5, 3.0, 3.5, 4.0\}$ and $\{1e-5, 1e-4, 1e-3, 1e-2, 1e-1, 1\}$. An AUC surface with respect to $(\gamma, -\text{lg }\lambda)$ is plotted in Fig. \ref{fig_a4_params}. It can be observed that if we fix $\lambda$ as a constant (e.g. $-\text{lg }\lambda = 2$), when $\gamma$ increases, the performance tends to climb up and remain stable. The larger the value of $-\text{lg }\lambda$, the flatter the trend of the variation. Similarly, if we set $\gamma$ to a fixed value, the AUC score increases and reaches a plateau area as $-\text{lg }\lambda$ gets larger. The best result is marked with a black triangle in Fig. \ref{fig_a4_params}, and the corresponding parameter pair $(\gamma, \lambda)$ reads $(2.0, 1e-4)$.

	\section{Conclusion}
	
	We focus on the deep learning-based HAD task, and address the issue that the reconstruction DNNs may learn an identical mapping between the input and output, which can lead to decreased detection performance. The main reason for this issue is that the traditional training treats the anomalies and background equally in a given HSI, making them hard to distinguish. In this paper, we propose a model-independent training strategy for DNNs that explicitly separates anomalies and background. A separation loss function based on a latent binary mask is proposed to reconstruct the background while suppressing anomalies. The efficient second-order LoG operator is used for anomaly suppression by alleviating large spatial variations. To maintain a high detection performance during training, the mask is periodically updated via an estimated proportion threshold. Experiments on benchmark datasets demonstrate the superiority of our method in HAD performance. Furthermore, the transferability of our training strategy is validated by applying it to several deep networks and achieving improved detection performance compared to the original networks.

	
	\bibliographystyle{IEEEtran}
        \input{ref/reference.tex}

	
	
	
	
	

	\vfill
	
\end{document}